\documentclass[aps,floatfix,twocolumn,prb,notitlepage,superscriptaddress,tightenlines,floatfix,10pt]{revtex4-2}

\usepackage{graphicx, color}
\usepackage{amsmath, amsthm, amssymb}
\usepackage{siunitx}
\usepackage{framed}
\usepackage{lipsum}  

\usepackage{hyperref}
\hypersetup{colorlinks,linkcolor=blue,urlcolor=blue,citecolor=blue}

\usepackage[version=4]{mhchem}
\usepackage{bm}
\usepackage{fixmath}
\usepackage{microtype}
\usepackage{hyperref}

\newcommand{\avg}[1]{\langle#1\rangle}

\def\a{\alpha}
\def\b{\beta}
\def\g{\gamma}

\def\e{\varepsilon}

\newcommand{\pd}{\partial}

\usepackage[titletoc]{appendix}

\begin{document}
\begin{abstract}
We study the continuum limit of two-dimensional chiral magnets in which Dzyaloshinskii-Moriya interaction (DMI) is due to the interplay between a smooth magnetic texture and spin-orbit coupling.
The resulting free-energy density of the system contains linear terms in the spatial gradient of the magnetic texture, which mark an instability of the system towards the formation of nontrivial magnetic orders such as skyrmions or chiral domain walls.
We perform a microscopic analysis of DMI tensors responsible for this contribution to free energy based on a Berry phase formulation in the mixed space of momentum and position, and reveal that they exhibit non-Lifshitz invariants features. 
In particular, a perturbation theory shows in the case of Rashba spin-orbit interactions the presence of non-Lifshitz invariants to third order in the small spin-orbit interaction and fourth order in the small exchange coupling.
The higher-order terms may even lead to an enhancement of DMI interaction at strong spin-orbit coupling due to divergences in the density of states at the bottom of the conduction band.
Finally, we also study the DMI free energy generated from Rashba spin-orbit interaction in different symmetry groups.
\end{abstract}
\title[Article]{Non-Lifshitz invariants corrections to Dzyaloshinskii-Moriya interaction energy}
\author{Doru Sticlet}
\email{doru.sticlet@itim-cj.ro}
\affiliation{National Institute for R\&D of Isotopic and Molecular Technologies, 67-103 Donat, 400293 Cluj-Napoca, Romania}
\author{Fr\'ed\'eric Pi\'echon}
\email{frederic.piechon@universite-paris-saclay.fr}
\affiliation{Universit\'e Paris-Saclay, CNRS, Laboratoire de Physique des Solides, 91405 Orsay, France}
\maketitle
	
\section{Introduction}
Chiral symmetry-breaking in magnetic materials results in an antisymmetric exchange coupling called the Dzyaloshinskii-Moriya interaction (DMI)~\cite{Dzyaloshinsky1958,Dzyaloshinskii1964,Moriya1960}, which tends to cant neighboring spins such that noncollinear magnetic orders are favored in the system~\cite{Bogdanov1989,Bogdanov2001,Roessler2006}.
As a consequence of DMI, nontrivial magnetic structures such as chiral domain walls~\cite{Thiaville2012, Emori2013, Ryu2013} and skyrmions~\cite{Muehlbauer2009, Yu2010, Jonietz2010,Adams2011,Heinze2011} become stable.
The latter are excitations in the form of magnetization vortices, which are topologically robust, and have been the subject of intense research in recent years~\cite{Soumyanarayanan2016, EverschorSitte2018, Bogdanov2020,Goebel2021}.
The controlled creation and annihilation of skyrmions with spin-polarized currents~\cite{Romming2013,Jiang2015}, gate voltages~\cite{Hsu2017,Schott2017}, or lasers~\cite{Berruto2018} feeds the driving goal to realize energy-efficient spintronic devices operating at room temperature for memory storage~\cite{Iwasaki2013,Fert2013,Tomasello2014,Yu2016}.  

A promising platform for probing such physics is in effectively two-dimensional systems where interfacial DMI develops~\cite{Hellman2017}.
In thin ferromagnetic films or in multilayers with alternating magnetic and nonmagnetic layers, the inversion symmetry is broken at the interfaces and thus a strong spin-orbit coupling (SOC) is generated.
In a long wavelength approach, the magnetic texture below the Curie temperature is described by the continuous magnetic density vector $\bm m(\bm r)$ of unit amplitude, with position $\bm r$ in the plane of the magnetic layer $(x,y)$.
The effect of SOC is to generate in the free energy linear terms in the texture gradient, which are characteristic for the DMI\@.
The micromagnetic DMI free energy follows from
\begin{equation}\label{om1}
	\Omega_{1} = 
	\sum_{\substack{\alpha\in\{x,y,z\} \\ j\in\{x,y\}}}w_{j\alpha} \frac{\pd m_\alpha}{\pd r_j}.
\end{equation}
The most common approach~\cite{Dzyaloshinskii1964} is to consider that energies $w_{j\alpha}$ are linear in magnetization $\bm m$,
\begin{equation}
	\Omega_1\simeq
	\frac12 D_{\alpha\beta,j} (m_\alpha\pd_{r_j} m_\beta -m_\beta\pd_{r_j} m_\alpha),
\end{equation}
which amounts to take into account only the well-known Lifshitz invariant (LI) contribution to $\Omega_1$ (see Appendix~\ref{sec:notation}).
The microscopic analytical calculation of DMI tensor $D_{\alpha\beta,j}$ in the continuum limit was only recently performed for the first time in topological insulators~\cite{Tserkovnyak2015,Wakatsuki2015} and a two-dimensional (2D) Rashba thin film~\cite{Ado2018}.
Notably, these were preceded by different approaches where DMI was explained in the vein of Ruderman-Kittel-Kasuya-Yosida theory as due to spin interactions mediated by conduction electrons~\cite{Imamura2004, Kundu2015}.
An analysis of effects beyond the Lifshitz invariant correction was performed in Ref.~\cite{Ado2020} in order to get a more general description of DMI in chiral magnets, and it has established that such corrections can be consequential. 
It was soon shown that indeed there are cases as in tetrahedral magnets where the conventional LI contribution vanishes by symmetry while the remaining non-LI contributions lead to a noncollinear magnetic structure~\cite{Ado2021,Rybakov2021}.

In this paper, we revisit the issue of non-Lifshitz invariants contribution to DMI from a different point of view, in which DMI is due to Berry curvature in phase space.
Our approach assumes that the magnetization $\bm m$ varies slowly in space on the scale of interatomic distance. 
Thus, the effect of $\bm m$ on the periodic Bloch wave functions is considered perturbatively.
In this sense, the Bloch wave vectors depend on position through $\bm m$, $|n, \bm k,\bm m(\bm r)\rangle$. 
In such cases it is natural to consider an approach based on a generalized Berry phase in the space defined by position and momentum~\cite{Xiao2005}.
Indeed, it was shown that the dynamics of electrons in the thin layer is determined by the Berry curvature in the phase space~\cite{Freimuth2013, Freimuth2014}.

Here we analyze the generic case of two-band systems with crossings near the $\mathit{\Gamma}$ point. In such cases it is analytically tractable to obtain the form of DMI tensors.
These are determined from the corresponding DMI energy $\Omega_1$, which follows from an expansion of the total grand-canonical thermodynamic free energy $\Omega$ in texture gradients.
This generates a contribution that is proportional to the Berry curvature in phase space~\cite{Freimuth2013}.
The expansion in gradient is also further refined with additional expansions in SOC or exchange amplitude.
This allowed us to determine the order at which non-LI contributions might become relevant.

As an application, the present study focuses on Rashba SOC, which occurs naturally in effective 2D systems due to the large variation in the electrostatic potential normal to the layer.
Usually the structure of the LI and non-LI invariants may be determined by symmetry analysis~\cite{Dzyaloshinskii1964,Bogdanov1989,Ado2020}.
Here we perform instead a microscopic analysis where such the structure is emergent from effective two-band Hamiltonians. 
Such models are based on the specific form of the Bloch bands at the $\mathit{\Gamma}$ point, as constrained by symmetries of the magnetic point groups. 
The form of Rashba SOC to cubic order in momentum was classified for 2D materials in Refs.~\cite{Samokhin2015a,Samokhin2022}, and constitutes for us a starting point in determining microscopically the DMI\@.

Our analysis reveals in the Berry curvature formulation of the problem, that the DMI free energy decomposes into two distinct parts $\Omega_1 = \Omega_1^{(0)} + \Omega_1^{(1)}$.
Usually only the first part has been a subject of investigation.
The second contribution, $\Omega_1^{(1)}$, is higher order in SOC, but nonetheless it is of the same order in the magnetization $\bm m$, and also enters to the same order in the exchange coupling strength.
For example, the second contribution in its lowest order in SOC is responsible for symmetric DMI tensors, which are usually discarded in the bulk, but may generate some nontrivial edge spin texture~\cite{Hals2017}.
The term $\Omega_1^{(1)}$ contains also antisymmetric DMI tensors, which renormalize the $\Omega_1^{(0)}$ contribution, and additionally, we show that they can lead to divergences in the free energy since they contain Fermi surface contributions, which diverge at low temperature due to singularities in the density of states.
All our investigations are made concrete in the study of effective models with Rashba SOC in different symmetry groups.

The article is organized as follows. 
Sec.~\ref{sec:model} introduces the class of two-band Hamiltonian models in which DMI develops.
The section also reviews the generic structure of the energy density $\Omega_1$ that is linear in a smooth spatial gradient of the magnetization, using a Berry phase formulation in the mixed space of momentum and position.
Sec.~\ref{sec:2-band} develops a perturbation theory, which uncovers the non-Lifshitz invariants corrections to $\Omega_1$.
The section expresses the form of generalized DMI tensors, and develops further expansions in the small and large SOC limit, relative to the exchange energy.
Sec.~\ref{sec:rashba} particularizes the analysis to the case of Rashba SOC in the $C_{\infty v}$ group.
Sec.~\ref{sec:sym} looks briefly at the DMI contribution from Rashba SOC in different symmetry groups.
Appendix~\ref{sec:appendix} details several of the points in the main paper such as: a determination of DMI constants for the conventional Rashba SOC in $C_{\infty v}$ group, an analysis of group $D_3$ where the SOC exhibits an out-of-plane component, a table with LI and $\Omega_1$ in all 10 2D groups obtained in the limit of small SOC, etc.
Sec.~\ref{sec:conclusion} summarizes the main points in the paper.

\section{Phase space Berry curvature formulation of spin-orbit-induced free-energy terms, linear in spatial magnetization gradient}\label{sec:model}

This section briefly recalls the derivation of the correction $\Omega_1$ to free-energy density that is linear in the gradient of the magnetic texture.
Starting from generic two-band Hamiltonian models, it is shown that the correction $\Omega_1$ writes as an average over occupied states of the momentum and position-dependent skyrmion-like density of a vector field $\bm h(\bm k,\bm r)$ that combines the spin-orbit coupling and the exchange coupling to the magnetic texture. Complementary to previous works, we show that this skyrmion density entails two distinct contributions that appear at different order in spin-orbit coupling, but nonetheless both contributing to same order in the magnetization $\bm m$ and exchange coupling.

\subsection{Model Hamiltonian.} In the following, we focus on generic two-band Hamiltonian models of the form
\begin{eqnarray}\label{ham}	
	H(\bm k,\bm r)&=& \xi(\bm k)\sigma_0+\bm h(\bm k,\bm r) \cdot\bm \sigma,\notag\\
	\bm h(\bm k,\bm r) &=& \Delta_{\rm so}\bm \gamma(\bm k) + \Delta_{sd} \bm m(\bm r),
\end{eqnarray}
with $\bm \sigma$ the vector of Pauli matrices, and $\sigma_0$ the identity matrix.
The first term is the energy dispersion of electrons in the absence of spin-orbit coupling, which is an even function of momentum $\xi(-\bm k)=\xi(\bm k)$.
The second contribution is a momentum- and position-dependent vector field $\bm h (\bm k,\bm r)$ that combines the spin-orbit coupling (SOC) and the exchange coupling to the magnetic texture. 
The SOC is described by an antisymmetric spin-orbit vector $\bm\gamma(-\bm k) = -\bm \gamma(\bm k)$ and a coupling strength $\Delta_{\rm so}$. 
The  magnetic exchange is characterized by a coupling strength $\Delta_{sd}=J_{sd}S$, with $S$, the magnitude of spins in the magnetic layer, and $J_{sd}$, the exchange coupling. 
The magnetic texture is modeled by a (unit length) vector $\bm m(\bm r)$ which varies smoothly in space.
As it appears below, the coupling strengths $\Delta_{\rm so}$ and $\Delta_{sd}$ are useful parameters to keep track of the order in a perturbation theory 
in weak spin-orbit or weak exchange coupling limits.

\subsection{Free-energy density.}
The free-energy density is obtained from the local density of states $\rho(\e,\bm r)$,
\begin{equation}
	\Omega(\bm r) = \int d\e \rho(\e,\bm r)g(\e),
\end{equation}
with $g(\e)$ the primitive of the Fermi-Dirac distribution function $f(\e) = g'(\e)$, $f(\e) = 1/(1+e^{\beta(\e-\mu)})$.
The local density of states is expressed using the Green's functions in a Wigner representation, in the mixed center-of-mass space coordinate $\bm r$ and relative momentum $\bm k$.
Assuming that the Green's functions vary slowly in space, it is advantageous to expand them in spatial gradients of the magnetization $\nabla_{\bm r}\bm m$. 
This translates in a gradient expansion of the density of states $\rho=\rho_0+\rho_1+\dots$, with the subscript denoting the order of the gradient (see Refs.~\cite{Freimuth2013, Freimuth2014, Ado2018, Brinker2019} and App.~\ref{sec:grad_expand} for details).

The effective density of states to linear order in the magnetization gradient reads
\begin{equation}\label{ldos}
	\rho(\e,{\bm r})=\langle(1-{\cal B}^{jj}_{s, \bm k} )\delta(\e -{\e}_{s,\bm k}-s\cdot h {\cal B}^{jj}_{s, \bm k})\rangle
\end{equation}
with the shorthand notation
\begin{equation}
	\avg{\ldots}
	\equiv  \sum_{s=\pm}\int \frac{d^d\bm k}{(2\pi)^d}\dots,
\end{equation}
and where summation over repeated indices $j$ is assumed.
In Eq.~\eqref{ldos}, $\e_{s,\bm k}$ is the semiclassical energy spectrum of the Hamiltonian in Eq.~\eqref{ham},
\begin{eqnarray}\label{eigs}
	\e_{s,\bm k}(\bm r) &=& \xi(\bm k)+s\cdot h(\bm k,\bm r),\notag\\
	h(\bm k, \bm r)&=&\sqrt{\Delta_{sd}^2+\Delta_{\rm so}^2\gamma^2
		+2\Delta_{\rm so}\Delta_{sd}\bm \gamma\cdot\bm m },
\end{eqnarray}
with $s=\pm$, the band index, and $h\equiv |\bm h|$.
Lastly,
\begin{equation}\label{berry}
	{\cal B}_{s, \bm k}^{ij}(\bm r)=-s\frac{1}{2}\frac{\bm h\cdot(\pd_{r_i}\bm h\times\pd_{k_j}\bm h)}{|\bm h|^3},
\end{equation}
denotes the element $(ij)$ of the intraband {\em phase space} Berry curvature tensor.

The expression~\eqref{ldos} illustrates two qualitatively distinct effects resulting from the gradient corrections. 
On the one hand, there is a momentum-position dependent shift of the band spectrum, and, on the other hand, there is also a modification of the spectral weight~\cite{Freimuth2013}. 
To linear order in the gradient, both effects are proportional to the phase-space Berry curvature Eq.~\eqref{berry}.
Using this effective density of states, the free-energy density is decomposed as $\Omega=\Omega_0+\Omega_1$,
with a zero-order contribution describing the uniform state, $\Omega_0({\bm r}) = \int d\e \rho_0(\e) g(\e)$, and a contribution $\Omega_1({\bm r})$, linear in the gradient of $\bm m$, which reads
\begin{eqnarray}
	\Omega_1({\bm r})& =
	&\avg{
		\frac{\bm h\cdot(\pd_{r_j}\bm h\times\pd_{k_j}\bm h)}{2}
		F_{s,\bm k}},
	\label{omega1}
\end{eqnarray}
with
\begin{equation}
	F_{s,\bm k}(\bm r) = \frac{
		sg(\e_{s,\bm k})- hf(\e_{s,\bm k})
	}{h^3}.
\end{equation}
Using the explicit expression of $\bm h$ from Eq.~\eqref{ham}, it follows that the correction to free-energy density $\Omega_1$ has two distinct contributions
\begin{equation}\label{decomp}
	\Omega_1^{} = \Omega_1^{(0)} + \Omega_1^{(1)},
\end{equation}
with
\begin{eqnarray}\label{om_1s}
	\Omega_1^{(0)}({\bm r}) &=& \frac{\Delta_{\rm so}\Delta^2_{sd}}{2}\avg{
		\pd_{k_j}\bm \gamma\cdot(\bm m\times\pd_{r_j}\bm m)
		F_{s,\bm k}
	}
	,\notag\\
	\Omega_1^{(1)}({\bm r}) &=& \frac{\Delta^2_{\rm so}\Delta_{sd}}{2}\avg{
		\pd_{r_j}\bm m\cdot(\pd_{k_j}\bm \gamma\times\bm \gamma)
		F_{s,\bm k}
	}
	.
\end{eqnarray}
These contributions $\Omega_1^{(0,1)}({\bm r})$ have a structure similar to the one in Eq.~\eqref{om1} and as detailed in the next section, 
both generate Lifshitz invariant and non-Lifshitz invariant contributions to the DMI interaction.

At this point, a few remarks are in order. 
The possibility to express the linear gradient corrections in Eqs.~\eqref{ldos} and~\eqref{omega1} solely in terms of the intraband phase-space Berry curvature is specific to two-band models. 
Likewise, the possibility to express the phase-space Berry curvature directly in terms of a phase-space {\em skyrmion-like density} 
of the vector field $\bm h(\bm k,\bm r)$ is also  specific to two-band models. However, Eqs.~\eqref{ldos} and~\eqref{omega1} are valid for any two-band model (in any dimension) of the form given by Eq.~\eqref{ham}.
Importantly, the expressions \eqref{omega1} and~\eqref{om_1s} contain full nonperturbative dependencies in the coupling strengths $\Delta_{\rm so}$ and $\Delta_{sd}$ 
and also full nonlinear dependencies in the magnetization vector $\bm m (\bm r)$ since all these parameters appear implicitly in $\bm h$ and $F_{s,\bm k}$.

\section{General expansion of non-Lifshitz invariant contributions}
\label{sec:2-band}
In the following, the free-energy density contributions $\Omega_1^{(0,1)}$ are expressed as Ginzburg-Landau-like expansions in $\bm m$ when considering $\Delta_{\rm so}\Delta_{sd}\bm \gamma\cdot\bm m/\lambda^2$ as a small parameter, with
\begin{equation}\label{lambda}
	\lambda = \sqrt{\Delta_{sd}^2+\Delta_{\rm so}^2\gamma^2}.
\end{equation}

Generically, the free-energy densities are expanded as
\begin{equation}
	\Omega_1^{(i)} = \sum_{n=0}^{\infty} \Omega^{(i)}_{1,n},
\end{equation}
with
\begin{equation}\label{dmi_intro}
	\Omega_{1,n}^{(i)} = D^{(i)}_{\alpha\beta\mu_1\dots\mu_{2n},j}
	(m_\alpha\pd_{r_j} m_\beta)m_{\mu_1}\!\cdots m_{\mu_{2n}},
\end{equation}
where $D^{(i)}$ are DMI tensors of odd rank.
The lowest-order term $n=0$ is quadratic in $\bm m$ and yields the Lifshitz invariant contributions $D^{(i)}_{\alpha\beta,j}$ of the DMI tensor.
The higher-order terms $n>0$ yield the non-Lifshitz-invariant contributions $D^{(i)}_{\alpha\beta\mu_1\dots\mu_{2n},j}$ of the DMI tensor. 
Since in higher-order contributions there is no requirement of antisymmetry in indices $\mu_1,\ldots \mu_{2n}$, these should be considered generalized DMI energies and tensors.

More quantitatively (see App.~\ref{sec:app_2} for details), the expansion of eigenenergies leads to an expansion of $F_{s,\bm k}$.
Note that since $\bm\gamma$ is antisymmetric in $\bm k$, only the symmetric part of $F_{s,\bm k}$ in $\bm k$ 
contributes to $\Omega_1^{(0)}$, and only the antisymmetric part of $F_{s,\bm k}$ contributes to $\Omega_1^{(1)}$,

\begin{eqnarray}\label{exp_a}
	\Omega_1^{(0)}&=&\sum_{n=0}^{\infty}
	\frac{\Delta_{\rm so}^{2n+1}\Delta_{sd}^{2n+2}}{2}\avg{\pd_{k_j}\bm \g\cdot(\bm m\times\pd_{r_j}\bm m) (\bm \g\cdot\bm m)^{2n}\notag\\
		&&{}\times\mathcal F_{s,\bm k}^{(2n)}(\lambda)},\\
	\Omega_1^{(1)}&=&\sum_{n=0}^{\infty}\frac{\Delta_{\rm so}^{2n+3}\Delta_{sd}^{2n+2}}{2}
	\avg{\pd_{r_j}\bm m\cdot (\pd_{k_j}\bm \g\times \bm \g)
		\mathcal (\bm \g\cdot\bm m)^{2n+1}\notag \\
		&&{}\times\mathcal F_{s,\bm k}^{(2n+1)}(\lambda)},\notag
\end{eqnarray}
where the coefficients $\mathcal F_{s,\bm k}^{(n)}$ are even in $\bm k$, and are determined iteratively 
\begin{equation}\label{f_coeff}
	\mathcal F^{(0)}_{s,\bm k}(\lambda)
	= F_{s,\bm k}\bigg|_{\bm \gamma\cdot\bm m=0},\;
	\mathcal F^{(n)}_{s,\bm k}(\lambda) = \frac{1}{n\lambda}
	\frac{\pd \mathcal F^{(n-1)}_{s,\bm k}(\lambda)}{\pd \lambda},
\end{equation}
for $n>0$.
The use of argument $\lambda$ in previous expressions implies that all dependence on energy $\e_{s,\bm k}$ simplifies to one on $\e_{s,\bm k}^{(0)}=\xi+s\lambda$.

The Eqs.~\eqref{dmi_intro} and \eqref{exp_a} readily yield the general form of DMI tensors
\begin{eqnarray}\label{dmi_tensors}
	D^{(0)}_{\alpha\beta\mu_1\dots\mu_{2n},j} &=& \frac{1}{2} 
	\Delta^{2n+1}_{\rm so}\Delta_{sd}^{2n+2}\epsilon_{\alpha\beta\delta}
	\avg{(\pd_{k_j}\g_\delta) \g_{\mu_1}\!\cdots \g_{\mu_{2n}} \notag\\
		&&{}\times\mathcal F_{s,\bm k}^{(2n)}(\lambda)},\label{dmso1}
	\\
	D^{(1)}_{\alpha\beta\mu_1\dots\mu_{2n},j} &=& \frac{1}{2} \Delta^{2n+3}_{\rm so}\Delta^{2n+2}_{sd}
	\epsilon_{\nu\beta\delta}
	\avg{\g_\alpha \g_\nu(\pd_{k_j}\g_{\delta}) \notag\\
		&&{}\times \g_{\mu_1}\!\cdots \g_{\mu_{2n}}
		\mathcal F_{s,\bm k}^{(2n+1)}(\lambda)},\label{dmso2}
\end{eqnarray}
with $\epsilon_{\alpha\beta\delta}$, the Levi-Civita symbol.

The usual Lifshitz invariants contribution to the energy is contained in $\Omega^{(i)}_{1,0}$, and the related DMI tensors are $D^{(i)}_{\alpha\beta,j}$.
The expansion beyond the first order is responsible for non-Lifshitz invariants.
Note that even to first order, there is a marked difference between the two tensors. 
The first tensor $D_{\alpha\beta,j}^{(0)}$ is antisymmetric in $\alpha$ and $\beta$ indices, while there is no such constraint on $D^{(1)}_{\alpha\beta,j}$. 
The symmetric part of the latter tensor is usually neglected since it multiplies a total derivative $\pd_{r_j}(m_\alpha m_\beta)$, and vanishes when integrating over the entire sample.
It was shown, however, that it has physical effects in generating specific magnetic textures at the sample boundary~\cite{Hals2017}. 
Since we treat here the case of an infinite system, we consider only the antisymmetric part. 

It is particularly revealing to truncate the free-energy density expansion to the first term where non-LI contributions are present. 
This is done either in the limit of small spin-orbit coupling, or small exchange coupling.
From Eqs.~\eqref{dmso1} and \eqref{dmso2}, it follows that at weak SOC the free energy is approximated
\begin{eqnarray}\label{om_so}
	\Omega_{1} &=& \Omega^{(0)}_{1,0}+\Omega^{(1)}_{1,0}+\Omega^{(0)}_{1,1}+\mathcal O(\Delta^5_{\rm so}/\Delta^5_{sd}),\\
	&\simeq& (D_{\alpha\beta,j}^{(0)}+D^{(1)}_{\alpha\beta,j}+D^{(0)}_{\alpha\beta\mu_1\mu_2,j}m_{\mu_1}m_{\mu_2})
	m_{\alpha}\pd_{r_j} m_\beta.\notag
\end{eqnarray}
Similarly, in the case of weak exchange coupling (or large SOC) $\Delta_{\rm so}\gg \Delta_{sd}$,
\begin{eqnarray}\label{om_sd}
	\Omega_{1} &=& \Omega^{(0)}_{1,0}+\Omega^{(1)}_{1,0}+ \Omega^{(0)}_{1,1}+\Omega^{(1)}_{1,1}
	+\mathcal O(\Delta_{sd}^6/\Delta_{\rm so}^6),
	\\
	&\simeq&\sum_{i=0,1}(D^{(i)}_{\alpha\beta,j} + D^{(i)}_{\alpha\beta\mu_1\mu_2,j}m_{\mu_1}m_{\mu_2})m_\alpha\pd_{r_j} m_\beta.\notag
\end{eqnarray}
Note that the power counting in the two expansions is different. At small SOC, the linear order in $\Delta_{\rm so}$ is contained in $\Omega_{1,0}^{(0)}$ alone. 
This contribution to free energy and all conventional LI invariants are therefore determined exactly in this limit from the analysis of $\Omega_{1,0}^{(0)}$.
In the Appendix~\ref{sec:LI} we have microscopically obtained the LI invariants in all 10 two-dimensional point groups by considering the symmetry-allowed spin-orbit coupling to cubic order in momentum.
In contrast, in the limit of large SOC or small exchange, both tensors $D^{(0,1)}_{\alpha\beta,j}$ already contribute at the lowest order $\Delta_{sd}^2$, such that both $\Omega_{1,0}^{(0)}$ and $\Omega_{1,0}^{(1)}$ are needed.
Finally, the explicit expression of $\mathcal F_{s,\bm k}^{(n)}$ coefficients~\eqref{f_coeff} up to $n=4$, necessary to give the dominant non-Lifshitz invariants in both limit cases of Eqs.~\eqref{om_so} and~\eqref{om_sd} are given in App.~\ref{sec:app_2}. 

\section{Application to Rashba spin-orbit coupling}\label{sec:rashba}

\begin{figure*}[t]
	\includegraphics[width=\textwidth]{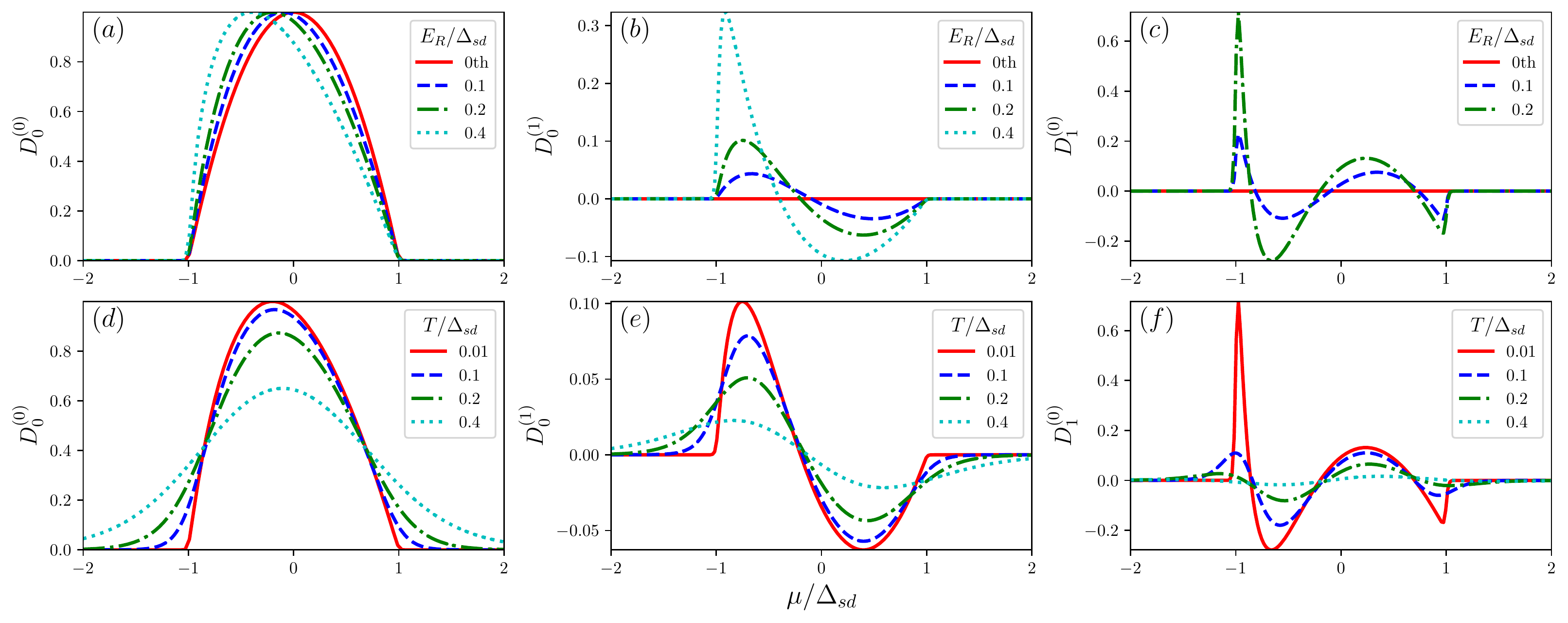}
	\caption{The DMI constants $D_0^{(0)}$, $D^{(1)}_0$, and $D^{(0)}_1$ in units of $k_R\Delta_{sd}/8\pi$ as a function of chemical potential in the limit of small spin-orbit coupling.
	Panels (a), (c), and (e) show the three DMI constants at different $E_R$ and $k_BT=0.01\Delta_{sd}$, with the red line represents denoting the zeroth-order approximation where the DMI constants are linear in $\alpha_R$. Panels (b), (d), and (f) present the same DMI constants' behavior at different temperatures and at fixed $E_R=0.2\Delta_{sd}$ [$k_B=1$].
	}\label{fig:cvinfd01d02}
\end{figure*}

The general theory from above is instantiated now in the important case of Rashba spin-orbit interactions.
The simplest case is that of the $C_{\infty v}$ group with a rotationally symmetric Rashba coupling $\Delta_{\rm so}\bm\gamma=\alpha_R(-k_y,k_x,0)$ for electrons with a parabolic spectrum,
\begin{equation}\label{rashba}
	H = (\frac{\hbar^2k^2}{2m}-\mu)\sigma_0+\alpha_R (\bm k\times\bm \sigma)_z+\Delta_{sd} \bm m \cdot \bm \sigma,
\end{equation}
with $\alpha_R$, the amplitude of Rashba SOC\@.

\subsection{Small SOC expansion.}
The limit of weak spin-orbit coupling relative to the exchange coupling $\Delta_{sd}$ is relevant in experiment and is the focus of the following.
To obtain the first non-LI invariant contribution to free-energy density it is necessary to expand $\Omega_1$ to cubic order in $\alpha_R$ as shown in Eq.~\eqref{om_so}.
That requires determining the tensors $D^{(0)}_{\alpha\beta,j}$, $D^{(1)}_{\alpha\beta,j}$, and $D^{(0)}_{\alpha\beta\mu_1\mu_2,j}$ (see App.~\ref{sec:cinfv} for details about the DMI tensors involved beyond the weak SOC approximation). 

Using the rotational symmetry of the Rashba SOC allows one to readily show that all nonzero tensor elements of $D^{(0)}_{\alpha\beta,j}$ are equal in amplitude, such that there is a single DMI constant characterizing the free-energy density
\begin{equation}
	\Omega^{(0)}_{1,0}=D^{(0)}_{0}
	L_{jz,j},
\end{equation}
with the DMI constant $D^{(0)}_0=D^{(0)}_{xz,x}$,
\begin{equation}\label{D1_0}
	D^{(0)}_0 = -\frac{\alpha_R\Delta_{sd}^2}{4\pi}\sum_s\int\frac{d k k}{\lambda^3}
	\Big(s g_{0,s}-\lambda f_{0,s}\Big),
\end{equation}
and Lifshitz invariant
\begin{equation}
	L_{\alpha\beta,j} = m_\alpha\pd_{r_j} m_\beta - m_\beta\pd_{r_j} m_\alpha.
\end{equation}
To first order in $\alpha_R$, $\lambda=\Delta_{sd}$ in Eq.~\eqref{lambda}, recovering the result in Ref.~\cite{Ado2018}.
The functions $f_{0,s}\equiv f(\e^{(0)}_{s,\bm k})$ and $g_{0,s}\equiv g(\e^{(0)}_{s,\bm k})$ are the Fermi-Dirac function and its primitive, respectively, evaluated in the zeroth-order approximation for the band energies $\e^{(0)}_{s,\bm k} = \xi+s\lambda$.

The tensors $D^{(1)}_{\alpha\beta,j}$, and $D^{(0)}_{\alpha\beta\mu_1\mu_2,j}$ are analyzed similarly, yielding the free energy contributions $\Omega^{(1)}_{1,0}$ and $\Omega_{1,1}^{(0)}$, respectively.
Since, again, in each tensor, the components are equal in amplitude, it is possible to factor out a single DMI constant in the free energies,
\begin{equation}
	\Omega_{1,0}^{(1)} = D_0^{(1)}L_{jz,j},\quad
	\Omega_{1,1}^{(0)} = D_1^{(0)}(1-m_z^2)L_{jz,j}.
\end{equation}
The constant $D_0^{(1)}=D_{xz,x}^{(1)}/2$ is obtained by extracting out the antisymmetric contribution in $D^{(1)}_{\alpha\beta,j}$. 
To cubic order in $\alpha_R$ reads
\begin{eqnarray}
	D_0^{(1)} &=& -\frac{\alpha_R^3\Delta^2_{sd}}{16\pi}\sum_s\int dk k^3 \mathcal F^{(1)}_{s,\bm k}(\Delta_{sd}).
\end{eqnarray}
Finally, the DMI constant $D_1^{(0)}=D_{xzxx,x}^{(0)}$ from $D^{(0)}_{\alpha\beta\mu_1\mu_2,j}$ has the expression to $\mathcal O(\alpha_R^3)$,
\begin{eqnarray}
	D^{(0)}_1 &=& -\frac{\alpha^3_R\Delta_{sd}^4}{8\pi}\sum_s\int dk k^3 \mathcal F^{(2)}_{s,\bm k}(\Delta_{sd}).
\end{eqnarray}
Therefore, the free-energy density $\Omega_1$ in this approximation is determined by all the three contributions
\begin{equation}
	\Omega_1 \simeq [D_0^{(0)}+D_0^{(1)}+(1 - m_z^2) D_1^{(0)}]
	L_{jz,j}
	.
\end{equation}
Already, to cubic order in $\alpha_R$ there are now non-LI invariants in the free energy $m_z^2L_{jz,j}$.
The additional dependence on $m_z^2$ is a 
property due to the rotational symmetry of the problem and was already predicted~\cite{Ado2020}.

Using natural momentum and energy scales characterizing the Rashba SOC, 
\begin{equation}
k_R=\frac{m\alpha_R}{\hbar^2}\text{ and } E_R=\frac{m\alpha_R^2}{2\hbar^2},
\end{equation}
respectively, yields simple analytical formulas for the constants in the zero-temperature approximation to $\mathcal O(\alpha_R^3)$,
\begin{align}
	D_0^{(0)} &\simeq \frac{k_R\Delta_{sd}}{8\pi}\big(
	1-\frac{2E_R\mu}{\Delta_{sd}^2}
	\big)
	\big(1-\frac{\mu^2}{\Delta_{sd}^2}\big)
	\Theta\big(1-\frac{\mu^2}{\Delta_{sd}^2}\big),\notag\\
	D_0^{(1)} &\simeq -\frac{k_R E_R \mu}{8\pi\Delta_{sd}}(1-\frac{\mu^2}{\Delta_{sd}^2})\Theta(1-\frac{\mu^2}{\Delta_{sd}^2}),\\
	D^{(0)}_1 &\simeq \frac{3k_RE_R\mu}{4\pi\Delta_{sd}}(1-\frac{5\mu^2}{3\Delta^2_{sd}})\Theta(1-\frac{\mu^2}{\Delta^2_{sd}}),
	\notag
\end{align}
with $\Theta$ the Heaviside function.
The results for $D_0^{(0)}$ from Ref.~\cite{Ado2018} are recovered by eliminating the cubic dependence on SOC by formally setting $E_R$ to $0$.
In the zero-temperature limit it follows that the DMI energy is nonvanishing only when the Fermi surface determined by $\mu$ is inside the exchange gap.
In this case there is a single circular Fermi surface at $k\simeq \sqrt{2m(\Delta_{sd}+\mu)}/\hbar$. Outside the exchange gap $\mu>\Delta_{sd}$, there are always two Fermi surfaces, with equal contribution and opposite sign, canceling in the sum over the bands. 

\begin{figure*}
	\includegraphics[width=\textwidth]{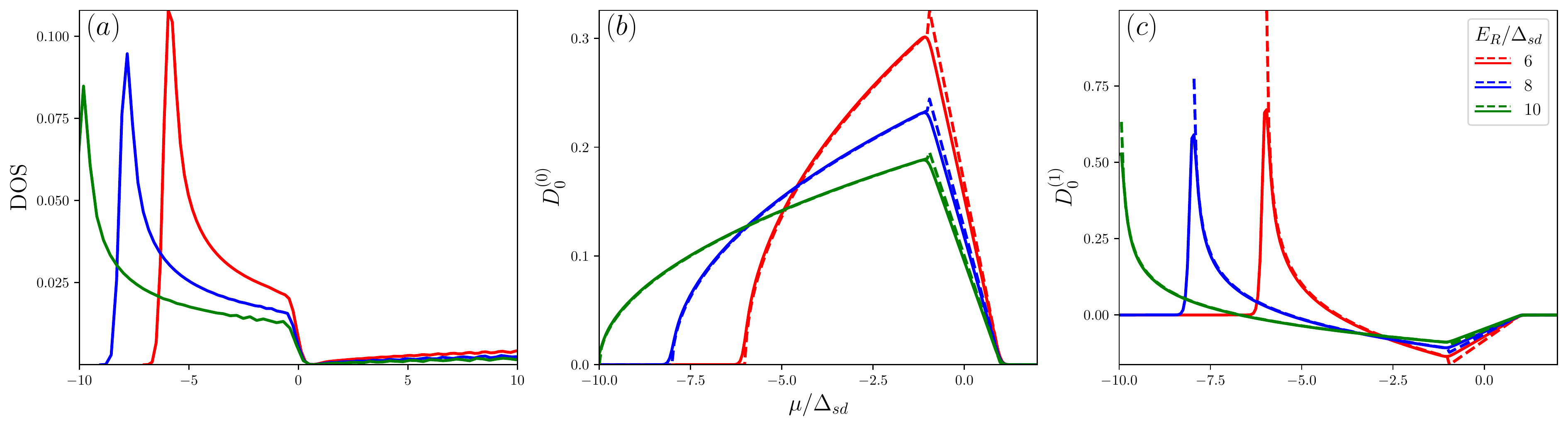}
	\caption{(a) Density of states at small $\Delta_{sd}$ (large SOC, $E_R/\Delta_{sd} > 1$) presents divergences at the bottom of the lowest band.
		Comparison between numerics (solid line) at $k_BT/\Delta_{sd}=0.1$ [$k_B=1$] and zero-temperature analytic approximation (dashed line) for (b) $D_0^{(0)}$ and (c) $D_0^{(1)}$ DMI constants in the Rashba $C_{\infty v}$ case in units of $k_R\Delta_{sd}/8\pi$.
	}
	\label{fig:cvinf_Delta}
\end{figure*}

The behavior of DMI constants for different relative strengths $E_R/\Delta_{sd}$ and at different temperatures are shown in Fig.~\ref{fig:cvinfd01d02}.
The zero-temperature approximation recovers the numerical behavior at low temperature and weak SOC $E_R/\Delta_{sd}\ll 1$.
At weak SOC the constant $D_0^{(0)}$ is symmetric in $\mu$, while $D_0^{(1)}$ and $D_1^{(0)}$, antisymmetric around the middle of exchange gap $\mu/\Delta_{sd}=0$.
Such symmetry is quickly lost at larger SOC and generally the constants have a higher value near the bottom of the gap, as explained below.
Larger corrections in $\alpha_R$ also lead to increasing the number of zeros in the free energy in their respective contribution at $\mu=0$.

With decreasing temperature and increasing $E_R/\Delta_{sd}$, the constants develop divergences at the bottom of the band. 
This is visible in Fig.~\ref{fig:cvinfd01d02}(b),~\ref{fig:cvinfd01d02}(c),~\ref{fig:cvinfd01d02}(e) and~\ref{fig:cvinfd01d02}(f) and it is due to the presence of derivatives of the Fermi-Dirac distribution in coefficients $\mathcal F_{s,\bm k}^{(1)}$ and $\mathcal F_{s,\bm k}^{(2)}$~\eqref{f_coeff_app}.
This effect cannot be captured analytically in the weak SOC expansion since $\alpha_R$ enters only as an overall prefactor, and the effective energy bands are determined by $\Delta_{sd}$ alone.
The divergence is, however, readily understood when considering $\alpha_R$ nonperturbatively.

\subsection{Large SOC expansion.}
It is telling to analyze this effect quantitatively in the opposite limit $E_R/\Delta_{sd}\gg 1$, although the effect is visible beyond this limit.
At large SOC, the two energy parabolas $\hbar^2k^2/2m$ for spin up and down are shifted, creating a degenerate manifold of momentum states with zero group velocity at $k\simeq k_R$ at the bottom of the lower band $\mu\simeq-E_R$.
Since the density of states is effectively one dimensional (1D) there, the total density of states will exhibit the usual inverse square-root energy singularity [Fig.~\ref{fig:cvinf_Delta}(a)]. 

More quantitatively, at zero temperature and in the limit of $E_R\gg\Delta_{sd}$, the leading approximation involves both $D_0^{(0)}$ and $D_0^{(1)}$ to $\mathcal O(\Delta_{sd}^2)$,
\begin{align}\label{large_soc}
	& D_0^{(0)}+D_0^{(1)}\simeq \frac{k_R\Delta_{sd}^2}{16\pi E_R} \notag\\
	& {}\times
	\begin{cases}
		(1+\frac{\mu}{E_R})^{\frac12}+(1+\frac{\mu}{E_R})^{-\frac12}, & \mu\in(-E_R,-\Delta_{sd}),\\
		1-\frac{\mu}{\Delta_{sd}},  &   \mu\in (-\Delta_{sd},\Delta_{sd}).
	\end{cases}
\end{align}
At $\mu=-\Delta_{sd}$, the two asymptotic expressions match to leading order in $\Delta_{sd}/E_R$. 
More importantly, near the band minimum at $\mu\simeq-E_R$, the constant $D^{(1)}_0$ displays the typical 1D singularity in the density of states $D^{(1)}_0\sim (1+\mu/E_R)^{-1/2}\sim 1/\sqrt{\e}$, where $\e$ is the energy calculated from $-E_R$.
The analytical results are corroborated with the numerical calculation of DMI constants presented in Fig.~\ref{fig:cvinf_Delta}(a) and~\ref{fig:cvinf_Delta}(c), where the typical  divergences in the 1D density of states are accompanied by the divergence in $D_0^{(1)}$.
Similar results are expected for higher-order terms in the expansion that contribute to order $\Delta_{sd}^4$ such as $D_1^{(0)}$ [see Fig.~\ref{fig:cvinfd01d02}(c) and~\ref{fig:cvinfd01d02}(f)] since they contain a stronger divergence generated by Fermi surface terms such as $f''(\e)$ that occur in $\mathcal F^{(3)}_{s,\bm k}$ in Eqs.~\eqref{f_coeff_app}.

The above considerations explain the divergences developing in DMI constants of higher order in $\alpha_R$ (see details in App.~\ref{sec:cinfv}).
This effect could be used as an exploit to single out non-LI contributions, with the provision that it would be seen only in the low-temperature regime, at strong SOC, with a chemical potential finely tuned near the lower band bottom.

\section{Rashba spin-orbit coupling in different symmetry groups}\label{sec:sym}
%
In order to analyze microscopically the DMI free energy in all 10 two-dimensional point groups, we consider effective SOC derived to cubic order in momentum in Ref.~\cite{Samokhin2022}.
Notably, in such cases, the rotational symmetry of $C_{\infty v}$ may be lost, and the SOC vector may develop out-of-plane components.
The latter is true in point groups where a $\pi$ rotation around $z$ axis is not a group element: $C_{1}$, $C_3$, $D_1$, and $D_3$.
In the remaining six groups, symmetry under a $\pi$ rotation and antisymmetry of $\gamma_z$, imposes $\gamma_z=0$. 
Consequently, in these groups there is a drastic reduction in the number of linearly independent components of the DMI tensors. 
Namely, from Eqs.~\eqref{dmso1} and \eqref{dmso2}, it follows that in the generalized DMI energy $m_z$ enters only once, and the generalized DMI tensors are reduced to $D^{(0,1)}_{izl_1\dots l_{2n},j}$, with Latin indices in the $(x,y)$ plane.

Let us briefly analyze the example of group $D_3$ in $\mathit{\Gamma}_4$ bands where the SOC vector develops an out-of-plane component, 
\begin{equation}\label{d3_soc}
	\Delta_{\rm so}\bm \gamma = (-\alpha_1 k_y,\alpha_1 k_x,\alpha_2 k_y(3k_x^2-k_y^2)).
\end{equation}
The SOC in this group is relevant for topological surface states of \ce{Bi2Te3} and \ce{Bi2Se3}~\cite{Fu2009,Liu2010}, \ce{BiTeI}~\cite{Bahramy2012}, hole gases in quasi-2D semiconductors~\cite{Moriya2014},
(001) surface states of oxide~\ce{SrTiO3}~\cite{Nakamura2012}, etc.

Note that the spin-orbit vector is identical in the $x$ and $y$ components to the case explored in the previous section, and therefore one expects to recover some of the same structure of DMI tensors from $C_{\infty v}$ case.
However, there is an additional cubic dependence on momentum in the $z$ component of the SOC vector.
The analysis in App.~\ref{sec:D3} shows that to first order in a perturbation theory $\Omega_{1,0}^{(0)}+\Omega_{1,0}^{(1)}$ there is no contribution from the cubic term, and the expected LI invariant follows, i.e.,~$L_{jz,j}$ generated by $\gamma_x$ and $\gamma_y$.
The effect of cubic Rashba term $\gamma_z$ is visible only at the level of non-LI invariants present in the material.
To cubic order in the SOC, there are now two non-LI invariants generated in the free-energy expansion.
One is identical to the previous  $C_{\infty v}$ case, and represents a quartic interaction of spins of the form $m_z^2L_{jz,j}$.
Additionally, there is a new invariant that involves only in-plane interactions between the spins,
\begin{equation}
	2m_x m_y L_{yx,x} + (m_x^2-m_y^2)L_{yx,y}.
\end{equation}
This non-LI is proportional (up to total derivatives that vanish in the bulk) to the invariant $m_x(m_x^2-3m_y^2)\partial_i m_i$ that was analyzed in detail in Ref.~\cite{Ado2022} for the group $D_{3h}$. 
The difference being that in our case this contribution to generalized DMI energy appears alongside the conventional LIs and the non-LI $m_z^2L_{jz,j}$.

The cubic terms in momentum in the SOC, such as those in $\gamma_z$ for $D_3$, are not reflected at the level of LI invariants, and generally may only contribute to higher-orders in the perturbation theory, to non-LI invariants.
To the fifth order in SOC, our calculations show that these terms only contribute to non-LI invariants only when the linear contribution in SOC coupling is present.
For the above case, that means the generalized DMI energy will contain to this order only terms of type $\alpha_1^n\alpha_2^m$ with $n>0$, (where $\alpha_1$ is the strength of the Rashba coupling linear in momentum).
Conversely, the Rashba coupling in $\mathit\Gamma_5$ and $\mathit\Gamma_6$ bands in $D_3$ group has no linear terms in momentum (see.~Tab.~\ref{tab:LI}), and yields no contribution to DMI energy to the lowest orders in spin-orbit coupling strength.

As a byproduct of the present theory, we also determine the conventional LI invariants, which follow in a first-order perturbation theory in weak SOC\@. 
A table of microscopically calculated DMI constants and LI-invariants in all symmetry groups is shown in Tab.~\ref{tab:LI} in App.~\ref{sec:LI}, and recovers the conventional invariants obtained in a standard symmetry analysis~\cite{Birss1966, Bogdanov1989}.


\section{Conclusions}\label{sec:conclusion}
In this article, we investigated generic two-dimensional, two-band continuum models where Dzyaloshinskii-Moriya interaction is generated in the interplay between spin-orbit coupling and a magnetic texture.
The DMI micromagnetic free energy, proportional to the first derivative in the gradient of a smooth magnetic texture, was analyzed in detailed to reveal its structure beyond the Lifshitz invariants corrections.
A second expansion in weak SOC or weak exchange coupling allows to pinpoint the exact order at which non-Lifshitz invariants are manifest, namely to third order in small SOC and fourth order in small exchange coupling.
The calculation of DMI tensors was performed in these limits explicitly for the case of rotation-symmetric $C_{\infty v}$ Rashba spin-orbit coupling.
A signature of higher-order terms is revealed in divergences in the generalized DMI energy due to singularities in the electronic bands. 
In the case of Rashba interactions this occurs due to the effective one-dimensional density of states near the bottom of the band at larger spin-orbit coupling, which generates an inverse square-root singularity in energy. 
Thus, a signature of non-Lifshitz invariants might be visible in measurements of the DMI constants, provided a strong SOC, a low temperature regime $k_BT\ll \Delta_{sd}$, with chemical potential tuned near the bottom of the band.

We have also shown how effective models for spin-orbit coupling in different point groups may be used to determine microscopically the generalized DMI energy. 
The lower symmetry of the Rashba vector compared to the continuum model with rotational symmetry induces new non-Lifshitz invariants. 
This approach is checked also by deducing the conventional LI invariants when taking only the first order in a weak SOC expansion.

A nontrivial extension to the present paper is the investigation of multiband effects in systems hosting skyrmions.
The free energy linear in the gradient of the magnetization is still expressed as a function of the Berry phase~\cite{Freimuth2013}, but a simple decomposition as in Eq.~\eqref{decomp} is not readily available.
Another open venue is the analysis of free-energy contributions that depend on higher-order gradients of the magnetization, which play a role in the stabilization of the skyrmion textures.  

\begin{acknowledgments} The authors thank A.~Thiaville for enlightening discussions on the topic.
	This work is supported by ``Investissements d'Avenir'' LabEx (ANR-10-LABX-0039-PALM).
	D.S.~also acknowledges financial support from the Romanian National Authority for Scientific Research and Innovation, UEFISCDI through the contract ERANET-QUANTERA QuCos 120/16.09.2019, and through Core Program 27N/03.01.2023, Project No.~PN 23 24 01 04.
\end{acknowledgments}

\newpage
\appendix
\onecolumngrid

\section{Gradient expansion}
\label{sec:grad_expand}
In this appendix, we detail the gradient expansion leading to the density of states approximation to linear order in the gradients from Eq.~\eqref{ldos}.
The calculation follows the lines drawn in Ref.~\cite{Freimuth2013} and is included to render the paper self-contained.

For an inhomogeneous system, the local density of states is obtained as $\rho(\e, {\bm r})= -\frac{1}{\pi}\text{Im}\text{Tr}[\mathcal G(\e,{\bm r},{\bm r})] $
where  $\mathcal G(\e,{\bm r},{\bm r'})$ is the (retarded) Green's functions, with the symbol $\text{Tr}$ corresponding to the trace over all internal degrees of freedom (spin/orbitals).
The Wigner representation of the Green's function is then defined as
\begin{equation}
	G(\e, \bm k, \bm r) = \int d \bm r' e^{-i\bm k\cdot\bm r'} 
	\mathcal G(\e,\bm r+\frac{\bm r'}{2},\bm r-\frac{\bm r'}{2}),
\end{equation}
with $\bm r$ playing the role of the center-of-mass position and $\bm k$ the relative momentum.
The local density of states then rewrites $\rho(\e, {\bm r})=-\frac{1}{\pi}\text{Im}\int \frac{d^d\bm k}{(2\pi)^d}\text{Tr}[G(\e, \bm k, \bm r)]$.
As explained in Ref.~\cite{Freimuth2013} the  Wigner Green's function $G(\e,{ \bm r}, {\bm k})$ is obtained from the Moyal product identity 
\begin{equation}\label{moyal}
	G_0^{-1}(\e,\bm k,\bm r) e^{\frac{i}{2}(\overleftarrow\nabla_{\bm r}\cdot \overrightarrow\nabla_{\bm k} 
		- \overleftarrow\nabla_{\bm k}\cdot \overrightarrow\nabla_{\bm r})} G(\e,\bm k,\bm r)=1,
\end{equation}
where $G_0^{-1} (\e, \bm k, \bm r)=\e -H({\bm k},{\bm r})$ with $H({\bm k},{\bm r})$ 
the Hamiltonian matrix [e.g., as given in Eq.~\eqref{ham}].
Expanding the Moyal identity to first order in gradients, and writing $G=G_0+G_1$, we obtain
\begin{equation}
	G_0^{-1}G_1^{} + \frac{i}{2}(\nabla_{\bm r}G_0^{-1}\cdot \nabla_{\bm k} G_0^{}-\nabla_{\bm k}G_0^{-1}\cdot \nabla_{\bm r} G_0^{})=0,
\end{equation}
where we use the identity $G_0^{-1}G_0^{}=1$.
Then the first correction $G_1$ reads as
\begin{equation}
	G_1 = \frac{i}{2}\sum_{j} G_0 [H_{r_j}G_0,H_{k_j} G_0],
\end{equation}
with $H_{r_j}=\partial_{r_j}H$,  $H_{k_j}=\partial_{k_j}H$ and where we used the identity $\partial_{r_j} G_0=G_0 H_{r_j} G_0$.
	
From now on, we focus on two-band models Hamiltonian of the form
$H({\bm k},{\bm r})=\xi(\bm k)\sigma_0+\bm h(\bm k,\bm r)\cdot\bm \sigma$ as given in Eq.~\eqref{ham}.
For this model, the zeroth-order Green's function writes as
\begin{equation}
	G_0(\e,\bm r,\bm k)=\sum_{s=\pm}\frac{P_s}{\e -\e_s+i\eta}, \text{ with } \e_{s}(\bm k,\bm r) = \xi+s h,\, P_s=\frac{1}{2}(1+s \frac{\bm h}{h}\cdot \bm \sigma), \text{ and } h\equiv|\bm h|.
\end{equation}
For brevity, the infinitesimal imaginary energy shift $\eta>0$ is neglected in the notation, but always implied in the following.
Correspondingly, the zeroth-order local density of states reads $\rho_0(\e,\bm r)=\int \frac{d^d\bm k}{(2\pi)^d} \sum_s \delta(\e-\e_s)$
where we use the identity $ -\frac{1}{\pi}\text{Im}\frac{1}{\e-\e_s}\equiv\delta(\e-\e_s)$.
Considering now the first-order gradient correction, it is convenient to define $g_1=\text{Tr}[G_1]$, which reads
\begin{equation}
	g_1(\e, \bm k,\bm r)=\sum_{s,j} {\cal B}_{s,\bm k}^{jj}\bigg[\frac{s
\cdot h}{(\e -\e_s)^2}-\frac{1}{\e-\e_s}\bigg],
\end{equation}
with ${\cal B}_{s,\bm k}^{ij}=-\frac{s}{2}\frac{{\bm h}\cdot ({\bm h}_{r_i} \times {\bm h}_{k_j})}{h^3}$
the $(ij)$ elements of the phase space Berry curvature tensor.
Then the first order gradient correction follows as
\begin{equation}
	\rho_1(\e,\bm r)=-\int \frac{d^d\bm k}{(2\pi)^d} \sum_{s,j} {\cal B}_{s,\bm k}^{jj}[s \cdot h \delta'(\e-\e_s)+\delta(\e-\e_s)],
\end{equation}
where $\delta'(\e-\e_s)=\partial_\e \delta(\e-\e_s)$.
It is then straightforward to verify that the local density expression $\rho(\e,\bm r)$, as given in Eq.~\eqref{ldos}, verifies   $\rho=\rho_0+\rho_1$ when expanded to first order in gradient corrections.

\section{Detailed DMI constant determination in Rashba models}\label{sec:appendix}
\subsection{Notation}\label{sec:notation}
The convention used in the article is that Greek letters denote indices that can take values in $\{x,y,z\}$, while Roman ones, only in the two-dimensional plane $\{x,y\}$ of the layer.
Einstein notation, where repeated indices are summed over, is also employed throughout the paper.
Spatial derivatives are denoted as $\pd_{r_j}$ and act in the 2D plane of the material.

The conventional generalized DMI tensor notation is related the one used in this paper as follows:
\begin{equation}
	D^{(i)}_{\alpha j\beta\mu_1\dots\mu_{2n}}\equiv D^{(i)}_{\alpha\beta\mu_1\dots\mu_{2n}, j}
\end{equation}
where the $j$ index is separated out since it corresponds in the free energy $\Omega_1$ to a spatial derivative $\pd_x$ or $\pd_y$ of the magnetization $\bm m$.

The free energy is expressed conveniently with the aid of LI invariants,
\begin{equation}\label{LI}
	L_{\alpha\beta,j} = m_\alpha\pd_{r_j} m_\beta - m_\beta\pd_{r_j} m_\alpha.
\end{equation}
Such invariants are also denoted in the literature as $L_{\alpha\beta}^{(j)}$.

\subsection{Free-energy expansion in two-dimensional, two-band models with spin-orbit interactions}\label{sec:app_2}
Here we present in more detail the model and the expansion of the free-energy density from Secs.~\ref{sec:model} and \ref{sec:2-band}.
To improve readability, some equations in the main text are restated.

The continuum model from the main text in Eq.~\eqref{ham},
\begin{equation}
	H = \xi(\bm k)\sigma_0+\bm h\cdot\bm \sigma,\quad 
	\bm h = \Delta_{\rm so}\bm \gamma(\bm k) + \Delta_{sd} \bm m(\bm r),
\end{equation}
with local energy eigenvalues
\begin{equation}
	\e_{s,\bm k}(\bm r) = \xi+s \sqrt{\Delta_{sd}^2+\Delta_{\rm so}^2\gamma^2+2\Delta_{\rm so}\Delta_{sd}\bm \gamma\cdot\bm m \bm(\bm r)}.
\end{equation}
The correction to the first contribution to the gradient expansion in the free-energy density $\Omega_1$ uses an expansion of energy eigenstates 
\begin{equation}
	\e_{s,\bm k} = \xi +s\lambda\sqrt{1+\eta},
	\quad
	\lambda = \sqrt{\Delta_{sd}^2+\Delta_{\rm so}^2\gamma^2},
	\quad
	\eta = \frac{2\Delta_{\rm so}\Delta_{sd}}{\lambda^2}\bm \gamma\cdot\bm m,
\end{equation}
with $\eta$ the small parameter. This procedure generates in the free-energy density a Ginzburg-Landau expansion in the magnetization $\bm m$.

The free-energy density in Eq.~\eqref{om_1s} are
\begin{eqnarray}
	\Omega_1^{(0)}({\bm r}) &=& \frac{\Delta_{\rm so}\Delta^2_{sd}}{2}\avg{
		\pd_{k_j}\bm \gamma\cdot(\bm m\times\pd_{r_j}\bm m)
		F_{s,\bm k}
	}
	,\notag\\
	\Omega_1^{(1)}({\bm r}) &=& \frac{\Delta^2_{\rm so}\Delta_{sd}}{2}\avg{
		\pd_{r_j}\bm m\cdot(\pd_{k_j}\bm \gamma\times\bm \gamma)
		F_{s,\bm k}
	}
	.
\end{eqnarray}
The functions $F_{s,\bm k}$ and implicitly the free-energy density $\Omega_1$ are expanded in powers of magnetization field
\begin{equation}
	F_{s,\bm k}=\sum_{n=0}^{\infty}\mathcal F^{(n)}_{s,\bm k}(\lambda)\Delta_{\rm so}^n\Delta_{sd}^n
	{(\bm \gamma\cdot\bm m)}^n,\quad	
	\Omega_1^{(i)} = \sum_{n=0}^{\infty} \Omega^{(i)}_{1,n}.
\end{equation}
The spin-orbit vector $\bm\gamma$ is antisymmetric in $\bm k$, while the expansion coefficients $\mathcal F$ are symmetric in $k$.
Therefore, only the symmetric in $\bm k$ part of $F_{s,\bm k}$  
contributes to $\Omega_1^{(0)}$, and only the antisymmetric part of $F_{s,\bm k}$, to $\Omega_1^{(1)}$, such that the following simplified expressions follow.
Each order in the expansion is related to the rank of a corresponding generalized DMI tensor $D$ in the following way:
\begin{equation}
	\Omega^{(i)}_{1,n} = D^{(i)}_{\alpha\beta\mu_1\dots\mu_{2n},j} (m_\alpha\pd_{r_j} m_\beta)m_{\mu_1}\!\cdots m_{\mu_{2n}},
\end{equation}
with generalized DMI tensors
\begin{eqnarray}
	D^{(0)}_{\alpha\beta\mu_1\dots\mu_{2n},j} &=& \frac{1}{2} 
	\Delta^{2n+1}_{\rm so}\Delta_{sd}^{2n+2}\epsilon_{\alpha\beta\delta}
	\avg{(\pd_{k_j}\g_\delta) \g_{\mu_1}\!\cdots \g_{\mu_{2n}} \mathcal F_{s,\bm k}^{(2n)}(\lambda)},
	\\
	D^{(1)}_{\alpha\beta\mu_1\dots\mu_{2n},j} &=& \frac{1}{2} \Delta^{2n+3}_{\rm so}\Delta^{2n+2}_{sd}
	\epsilon_{\nu\beta\delta}
	\avg{\g_\alpha \g_\nu(\pd_{k_j}\g_{\delta}) \g_{\mu_1}\!\cdots \g_{\mu_{2n}} \mathcal F_{s,\bm k}^{(2n+1)}(\lambda)}.
\end{eqnarray}
In calculations, it is profitable to perform a decomposition of the tensors in symmetric and antisymmetric parts, thus revealing a reduction in the number of linearly independent components,
\begin{eqnarray}
	D^{(0)}_{\alpha\beta\mu_1\dots\mu_{2n},j}&=&D^{(0)}_{[\alpha\beta](\mu_1\dots\mu_{2n}),j},\\
	D^{(1)}_{\alpha\beta\mu_1\dots\mu_{2n},j}&=&D^{(1)}_{[\alpha\beta](\mu_1\dots\mu_{2n}),j} + 
	D^{(1)}_{(\alpha\beta)(\mu_1\dots\mu_{2n}),j}.\notag
\end{eqnarray}
Here $[\dots]$ and $(\dots)$ denote antisymmetric and symmetric tensor in those indices, respectively.
Further reductions are apparent only by considering specific point groups under which DMI tensors transform.

For practical purposes the free-energy density may be analyzed analytically in the weak SOC or weak exchange coupling limits to identify the leading non-LI contributions.
This leads to the truncated expansions in the $\Delta_{\rm so}\ll \Delta_{sd}$ limit,
\begin{eqnarray}
	\Omega_{1} &=& \Omega^{(0)}_{1,0}+\Omega^{(1)}_{1,0}+\Omega^{(0)}_{1,1}+\mathcal O(\Delta^5_{\rm so}/\Delta^5_{sd})\simeq (D_{\alpha\beta,j}^{(0)}+D^{(1)}_{\alpha\beta,j}+D^{(0)}_{\alpha\beta\mu_1\mu_2,j}m_{\mu_1}m_{\mu_2})
	m_{\alpha}\pd_{r_j} m_\beta,\notag
\end{eqnarray}
or $\Delta_{\rm so}\gg \Delta_{sd}$ limit,
\begin{eqnarray}
	\Omega_{1} &=& \Omega^{(0)}_{1,0}+\Omega^{(1)}_{1,0}+ \Omega^{(0)}_{1,1}+\Omega^{(1)}_{1,1}
	+\mathcal O(\Delta_{sd}^6/\Delta_{\rm so}^6)\simeq\sum_{i=0,1}(D^{(i)}_{\alpha\beta,j} + D^{(i)}_{\alpha\beta\mu_1\mu_2,j}m_{\mu_1}m_{\mu_2})m_\alpha\pd_{r_j} m_\beta.\notag
\end{eqnarray}

Computing the first terms in the free-energy density expansions above [or Eqs.~\eqref{om_so} and~\eqref{om_sd} in the main text] to obtain the non-LI invariants requires the first four coefficients determined from Eq.~\eqref{f_coeff}:
\begin{eqnarray}\label{f_coeff_app}
	\mathcal F^{(0)}_{s,\bm k}(\lambda)
	&=& \frac{1}{\lambda^3}(s g_{0,s}-\lambda f_{0,s})\notag\\
	\mathcal F^{(1)}_{s,\bm k}(\lambda)
	&=& \frac{1}{\lambda^5}(-3sg_{0,s}+3\lambda f_{0,s}-s\lambda^2 f'_{0,s}),\\
	\mathcal F^{(2)}_{s,\bm k}(\lambda)
	&=& \frac{1}{2\lambda^7}(15sg_{0,s} - 15\lambda f_{0,s}+6s\lambda^2f_{0,s}'-\lambda^3 f_{0,s}''),
	\notag\\
	\mathcal F^{(3)}_{s,\bm k}(\lambda)
	&=&\frac{1}{6\lambda^9}(-105sg_{0,s}+105\lambda f_{0,s}-45s\lambda^2 f_{0,s}'+10\lambda^3 f_{0,s}''-s\lambda^4 f_{0,s}''').\notag
\end{eqnarray}
The primes denote derivatives with respect to the energy argument of the Fermi-Dirac functions. 
Also, $g_{0,s}\equiv g(\e^{(0)}_{s,\bm k})$, $f_{0,s}\equiv f(\e^{(0)}_{s,\bm k})$, and derivatives are evaluated at $\e^{(0)}_{s,\bm k} = \xi+s\lambda$ at vanishing $\bm \gamma\cdot\bm m$.
Higher-order coefficients $\mathcal F$ contain derivatives of the Fermi-Dirac distribution function, which capture mainly Fermi surface contribution to the DMI tensor elements. 
Although such terms are small in a perturbation theory in either small SOC or small exchange, they can yield divergences in the free energy at small temperature when the density of states diverges such as for flat bands, van Hove singularities etc.

\subsection{\texorpdfstring{Group $C_{\infty v}$}{Cinfv}}\label{sec:cinfv}
This subsection details the calculation of generalized DMI tensors and constants in the Rashba model from Eq.~\eqref{rashba} in group $C_{\infty v}$. 
The spin-orbit coupling vector in this group is given by
\begin{equation}
	\Delta_{\rm so}\bm \gamma = \alpha_R(-k_y,k_x,0),
\end{equation}
and it is identical to the spin-orbit coupling in $D_4$: $\mathit{\Gamma}_6$ and $\mathit{\Gamma}_7$, and $D_6$: $\mathit{\Gamma}_7$ and $\mathit{\Gamma}_8$.
The $\bm \gamma$ expression determines the DMI tensor elements when using Eq.~\eqref{dmi_tensors}. We derive in the following the general form of the first four DMI tensors which capture the dominant contribution to non-LI invariants. 
Later in the subsection we perform a perturbation theory in either weak SOC or weak exchange coupling to get explicit forms for the DMI constants.

The DMI tensor in $\Omega^{(0)}_{1,0}$ is sparse with only four nonzero elements, which are equal in amplitude,
\begin{equation}
	D^{(0)}_{jz,j} = -D^{(0)}_{zj,j} = 
	-\frac{\alpha_R\Delta_{sd}^2}{2}\avg{\mathcal F_{s,\bm k}^{(0)}(\lambda) }.
\end{equation}
In these tensor elements the repeated indices are not summed. This convention also applies below and in the next sections whenever discussing a given DMI tensor element.
Factoring out one of the elements determines the DMI constant
\begin{equation}
	\Omega_{1,0}^{(0)} = D^{(0)}_0 (m_j\pd_{r_j} m_z - m_z\pd_{r_j} m_j) = D^{(0)}_0 L_{jz,j}
\end{equation}
with $D^{(0)}_0 = D^{(0)}_{xz,x}$ or, explicitly
\begin{equation}\label{D01}
	D_0^{(0)}= -\frac{\alpha_R\Delta_{sd}^2}{4\pi}\sum_s\int d k k
	\mathcal F^{(0)}_{s,\bm k}(\lambda).
\end{equation}

The next contribution is from the three-rank tensor $D^{(1)}_{\alpha\beta,j}$. The nontrivial tensor elements are
\begin{eqnarray}
	D^{(1)}_{xz,x} &=& -\frac{\alpha_R^3\Delta_{sd}^2}{2}\avg{k_y^2\mathcal F^{(1)}_{s, k}(\lambda)},\quad
	D^{(1)}_{yz,y} = -\frac{\alpha_R^3\Delta_{sd}^2}{2}\avg{k_x^2\mathcal F^{(1)}_{s, k}(\lambda)},\notag\\
	D^{(1)}_{xz,y} &=&D^{(1)}_{yz,x} = \frac{\alpha_R^3\Delta_{sd}^2}{2}\avg{k_x k_y \mathcal F^{(1)}_{s, k}(\lambda) }.
\end{eqnarray}

The last two elements vanish by using the spherical symmetry of the problem in the integrals over momentum in $\avg{\dots}$, and the remaining tensor elements read
\begin{equation}\label{D20_so}
	D^{(1)}_{xz,x} = D^{(1)}_{yz,y} = -\frac{\alpha_R^3\Delta_{sd}^2}{4}\avg{k^2\mathcal F^{(1)}_{s,\bm k}(\lambda)},\quad k^2=k_x^2+k_y^2.
\end{equation}
The symmetric part of the tensor integrates to zero over the bulk as it multiplies a total derivative $\pd_{r_j}(m_j m_z)$. Therefore, the nonvanishing part of the free-energy density has only the antisymmetric part 
\begin{equation}
	\Omega_{1,0}^{(1)} = D^{(1)}_0 
	L_{jz,j}
	,\quad D_0^{(1)} = D^{(1)}_{xz,x}/2,
\end{equation}
with explicit DMI constant
\begin{equation}\label{D02}
	D_0^{(1)}=-\frac{\alpha_R^3\Delta_{sd}^2}{16\pi}\sum_s\int dkk^3
	\mathcal F^{(1)}_{s,\bm k}(\lambda).
\end{equation}

The DMI tensor in $\Omega_{1,1}^{(0)}$ has eight nonvanishing tensor components,
\begin{equation}
	D^{(0)}_{jzii,j}=-D^{(0)}_{zjii,j}
	= -\frac{\alpha_R^3\Delta_{sd}^4}{4}\avg{k^2\mathcal F_{s,\bm k}^{(2)}(\lambda)}.
\end{equation}
Therefore, the free-energy correction reads
\begin{equation}
	\Omega_{1,1}^{(0)}=D^{(0)}_1(
	m_x\pd_x m_z - m_z\pd_x m_x)(m_x^2+m_y^2) + (x\leftrightarrow y) = D^{(0)}_1 (1-m_z^2)
	L_{jz,j},
\end{equation}
using $m^2 = 1$ in the second equality, with DMI constant $D^{(0)}_1=D^{(0)}_{xzxx,x}$, or
\begin{equation}\label{D11}
	D_1^{(0)}=-\frac{\alpha_R^3\Delta_{sd}^4}{8\pi}\sum_s\int dk k^3
	\mathcal F^{(2)}_{s,\bm k}(\lambda).
\end{equation}

The final generalized DMI tensor considered here has 16 non-vanishing components (not explicit here), leading to a free-energy density contribution,
\begin{equation}
	\Omega_{1,1}^{(1)} = 2D_1^{(1)}(1-m_z^2)(m_x\pd_x m_z + m_y\pd_y m_z).
\end{equation}
The free-energy density, after eliminating total derivatives $\pd_{r_j}(m_j m_z)$ and $\pd_{r_j}(m_j m_z^3)$, also reads
\begin{equation}
	\Omega_{1,1}^{(1)} = D_1^{(1)}
	(1-\frac{m_z^2}{2})
	L_{jz,j}
	,
\end{equation}
which presents the same $m_z^2$ correction to Lifshitz invariants. 
The DMI constant reads
\begin{equation}\label{D12}
	D_1^{(1)} = -\frac{3\alpha_R^5\Delta_{sd}^4}{64\pi}\sum_s\int dk k^5 \mathcal F_{s,\bm k}^{(3)}(\lambda).
\end{equation}

In the following we introduce the characteristic Rashba momentum and energy scales
\begin{equation}
	k_R = \frac{m\alpha_R}{\hbar^2},
	\quad
	E_R = \frac{m\alpha^2_R}{2\hbar^2},
\end{equation}
and perform a perturbation theory either in the small or large SOC limit.

\paragraph{Small SOC.\\}
We consider now the expansion in $\alpha_R/\Delta_{sd}$ as in Eq.~\eqref{om_so}.
%
%
In Fig.~\ref{fig:cvinfd01d02} we have shown the exact behavior of the DMI constants by numerical integration over bands and momentum, including the limit of small SOC.
Analytically, we also compute the DMI constants in the zero temperature limit, to the lowest orders in $\alpha_R$.

To $\mathcal O(\alpha_R^3)$, the zero-temperature DMI constants read as follows:
\begin{equation}\label{D01_so}
	D_0^{(0)} \simeq \frac{k_R\Delta_{sd}}{8\pi}\bigg(
	1-\frac{2E_R\mu}{\Delta_{sd}^2}
	\bigg)
	\bigg(1-\frac{\mu^2}{\Delta_{sd}^2}\bigg)
	\Theta\bigg(1-\frac{\mu^2}{\Delta_{sd}^2}\bigg),
\end{equation}
\begin{equation}
	D_0^{(1)} \simeq -\frac{k_R E_R \mu}{8\pi\Delta_{sd}}\bigg(1-\frac{\mu^2}{\Delta_{sd}^2}\bigg)
	\Theta\bigg(1-\frac{\mu^2}{\Delta_{sd}^2}\bigg),
\end{equation}
and
\begin{equation}
	D^{(0)}_1 \simeq  \frac{3k_R E_R\mu}{4\pi\Delta_{sd}}\bigg(1-\frac{5\mu^2}{3\Delta^2_{sd}}\bigg)
	\Theta\bigg(1-\frac{\mu^2}{\Delta^2_{sd}}\bigg),
\end{equation}
with Heaviside step function $\Theta$. To obtain $D_0^{(1)}$ it was necessary to expand $F_{s,\bm k}^{(0)}(\lambda)$ from Eq.~\eqref{D01} to $\alpha^3_R$, hence the term proportional to $\alpha_R E_R$. 
Note that to linear order in $\alpha_R$, only $D_0^{(1)}$ survives by formally setting in the expression $E_R=0$, such that it reproduces the results in Ref.~\cite{Ado2018}.
The zero-temperature results are obtained by performing the sums and integrals in Eqs.~\eqref{D01}, \eqref{D02}, and \eqref{D12} using Fermi-Dirac formulas at zero temperature $g(\e)=(\e-\mu)\Theta(\mu-\e)$, $f(\e) = \Theta(\mu-\e)$, and $f'(\e)=-\delta(\mu-\e)$.

\paragraph{Large SOC.}\mbox{}

In the limit of $E_R\gg\Delta_{sd}$ we obtain from Eq.~\eqref{D01} the leading zero-temperature approximation to $D_0^{(0)}$,
\begin{equation}
	D_0^{(0)}\simeq \frac{k_R\Delta_{sd}^2}{8\pi E_R}\times
	\begin{cases}
		2\sqrt{1+\mu/E_R}, & \mu\in(-E_R,-\Delta_{sd}),          \\
		1-\mu/\Delta_{sd}, & \mu\in (-\Delta_{sd},\Delta_{sd}).
	\end{cases}
\end{equation}
To same order in $\Delta_{sd}$ there is the additional contribution from $D_0^{(1)}$
from Eq.~\eqref{D02},
\begin{equation}
	D_0^{(1)}=\frac{k_R\Delta_{sd}^2}{16\pi E_R}\times
	\begin{cases}
		-3(1+\mu/E_R)^{1/2}+(1+\mu/E_R)^{-1/2}, & \mu\in(-E_R,-\Delta_{sd}),  \\
		\mu/\Delta_{sd}-1,              & \mu\in (-\Delta_{sd},\Delta_{sd}).
	\end{cases}
\end{equation}
Therefore, the DMI constant at large SOC (small exchange coupling)  $D_0^{(0)}+D_0^{(1)}$  is the one in Eq.~\eqref{large_soc}.
The $D^{(1)}_0$ constant has a divergence at the bottom of the band in the limit of small exchange coupling (or large SOC) due to the Fermi surface contribution to the free-energy density.
This is a consequence of the large density of states that develops at the bottom of the band, where the minimum occurs on a circle of constant energy at $k=k_R$, such that the density of states there has a characteristic divergence of a 1D model. 
The DMI constant has a similar divergence near the band minimum at $\mu\simeq-E_R$,
\begin{equation}
	D_0^{(1)} \sim (1+\mu/E_R)^{-1/2}\sim 1/\sqrt{\e},
\end{equation}
where $\e$ is the energy calculated from $-E_R$. 
Such effects start to be visible at low temperature even at $E_R<\Delta_{sd}$ in Fig.~\ref{fig:cvinfd01d02}(b) and (e), and more so at $E_R>\Delta_{sd}$, where the divergences in the DOS are accompanied by the divergence in $D_0^{(1)}$, respectively in Figs.~\ref{fig:cvinf_Delta}(a) and~\ref{fig:cvinf_Delta}(c). 
This situation becomes more visible for higher terms in the expansion that contribute to order $\Delta_{sd}^4$ such as $D_1^{(0)}$ [see Figs.~\ref{fig:cvinfd01d02}(c) and~\ref{fig:cvinfd01d02}(f)] since they contain a stronger divergence generated by Fermi-surface terms such as $f''(\e)$ that occur in $F^{(3)}_{s,\bm k}$ in Eqs.~\eqref{f_coeff}.

\subsection{Group \texorpdfstring{$D_3$}{ D3}}
\label{sec:D3}
There are cases where the SOC vector develops nonzero out-of-plane components where one could expect qualitatively different results.
This occurs for groups where rotation symmetry by $\pi$ is absent: $C_{1,3}$ and $D_{1,3}$.
This subsection details the calculation of generalized DMI tensors and constants in the Rashba model from Sec.~\ref{sec:sym} in group $D_3$ for $\mathit{\Gamma}_4$ bands. 
In this case, the spin-orbit vector reads
\begin{equation}
	\Delta_{\rm so}\bm \gamma = (-\alpha_1 k_y,\alpha_1 k_x,\alpha_2 k_y(3k_x^2-k_y^2)),
\end{equation}
where for convenience $\alpha_2$ is defined as half of $\alpha_2$ from Tab.~\ref{tab:LI}.
The cubic term breaks the rotational symmetry of the spectrum.

In the following, we will analyze the first terms, $\Omega_{1,0}^{(i)}$ and $\Omega_{1,1}^{(i)}$, in the Ginzburg-Landau expansion in magnetization $\bm m$ without assuming either relative small SOC, or small exchange coupling. At the end of the section, the small SOC will be treated in more detail since it allows analytical resolution for the DMI constants.

The non-vanishing components of $D^{(0)}_{\alpha\beta,j}$ tensor read
\begin{eqnarray}\label{D3_D1_sd}
	D^{(0)}_{jz,j}&=& -D^{(0)}_{zj,j} = -\frac{\alpha_1\Delta_{sd}^2}{2}
	\avg{\mathcal F_{s,\bm k}^{(0)}(\lambda)},\notag\\
	D^{(0)}_{xy,x}&=& -D^{(0)}_{yx,x}=\frac{6\alpha_2\Delta_{sd}^2}{2}
	\avg{k_xk_y\mathcal F_{s,\bm k}^{(0)}(\lambda)},\\
	D^{(0)}_{xy,y}&=& -D^{(0)}_{yx,y}=
	\frac{3\alpha_2\Delta_{sd}^2}{2}\avg{(k_x^2-k_y^2)\mathcal F_{s,\bm k}^{(0)}(\lambda)},\notag
\end{eqnarray}
with
\begin{equation}
	\lambda = \sqrt{\Delta_{sd}^2+\alpha_1^2k^2+\alpha_2^2k_y^2(3k_x^2-k_y^2)^2}.
\end{equation}
The last two equations in~\eqref{D3_D1_sd} vanish by symmetry and therefore the free energy contribution reads
\begin{equation}
	\Omega_{1,0}^{(0)} = -\frac{\alpha_1\Delta_{sd}^2}{2}\sum_s\int \frac{d^2\bm k}{4\pi^2}
	\mathcal F_{s,\bm k}^{(0)}(\lambda)L_{jz,j}.
\end{equation}
It exhibits the usual LI structure (see Tab.~\ref{tab:LI}).

There are eight components of the $D^{(1)}_{\alpha\beta,j}$ tensor that do not vanish under the constraint $D_3$ group imposes on the angular integral,
\begin{eqnarray}
	D_{iz,i}^{(1)} &=& -\frac{\alpha_1^3\Delta_{sd}^2}{2}\avg{k_{\bar i}^2\mathcal F^{(1)}_{s,\bm k}(\lambda)},\quad
	D_{xx,x}^{(1)} = \frac{\alpha_1^2\alpha_2\Delta_{sd}^2}{2}\avg{k_y^2(3k_x^2+k_y^2)\mathcal F^{(1)}_{s,\bm k}(\lambda)},\quad
	D_{xy,y}^{(1)} = -\alpha_1^2\alpha_2\Delta_{sd}^2\avg{k_y^4\mathcal F^{(1)}_{s,\bm k}(\lambda)},\notag\\
	D_{yy,x}^{(1)} &=& -3\alpha_1^2\alpha_2\Delta_{sd}^2\avg{k_x^2k_y^2\mathcal F^{(1)}_{s,\bm k}(\lambda)},\quad
	D_{yx,y}^{(1)} = \frac{3\alpha_1^2\alpha_2\Delta_{sd}^2}{2}\avg{k_x^2(k_y^2-k_x^2)\mathcal F^{(1)}_{s,\bm k}(\lambda)},\\
	D_{zx,x}^{(1)} &=& \frac{\alpha_1\alpha_2^2\Delta_{sd}^2}{2}\avg{k_y^2(k_y^4-9k_x^4)\mathcal F^{(1)}_{s,\bm k}(\lambda)},
	\quad
	D_{zy,y}^{(1)} = \alpha_1\alpha_2^2\Delta_{sd}^2\avg{k_y^4(3k_x^2-k_y^2)\mathcal F^{(1)}_{s,\bm k}(\lambda)}.
	\notag
\end{eqnarray}
Due to the lack of rotation symmetry it is not immediate to resolve these integrals as was the case in $C_{\infty v}$.
Using polar coordinates and  adding the contribution from all tensors as in Eq.~\eqref{dmi_intro} yields the energy density,
\begin{eqnarray}
	\Omega^{(1)}_{1,0} &=& - \frac{\Delta_{sd}^2}{2}\int \frac{dk k}{4\pi^2} 
	\bigg[
	\frac{\alpha_1^3 k^2}{2}(m_x\pd_x m_z+m_y\pd_y m_z)r_0^{(1)}(k)\\
	&&{}
	+\frac{3\alpha_1^2\alpha_2k^4}{4}(m_y\pd_x m_y-m_x\pd_x m_x+m_y\pd_y m_x+m_x\pd_y m_y)r_0^{(1)}(k)
	+\frac{\alpha_1\alpha_2^2k^6}{2}(m_z\pd_x m_x +m_z\pd_y m_y)r_1^{(1)}(k)
	\bigg],\notag
\end{eqnarray}
where the angular integral acts inside functions $r_{0,1}^{(1)}$. These are generally defined for following use,
\begin{equation}\label{r_func}
	r^{(m)}_n(k) = \sum_s\int_0^{2\pi} d\theta \sin(3\theta)^{2n}\mathcal F^{(m)}_{s,\bm k}(\lambda).
\end{equation}
The second term $\mathcal O(\alpha_1^2\alpha_2)$ vanishes since it contains only total derivatives over products of magnetization components.
Then after factoring out the symmetric part of the rest of the components, which integrates to zero in the bulk, one obtains
\begin{eqnarray}\label{d3_om2_10}
	\Omega^{(1)}_{1,0} &=& D_0^{(1)}L_{jz,j},\quad D_0^{(1)}=- \frac{\alpha_1\Delta_{sd}^2}{32\pi^2}\int dk k^3[\alpha_1^2r_0^{(1)}(k)-k^4\alpha_2^2 r_1^{(1)}(k)].
\end{eqnarray}
Thus, the usual LI invariant is indeed recovered to this order and a unique DMI constant is defined.

Higher-order tensors are expected to yield the non-LI contributions.
There are 28 nonvanishing components to tensors $D^{(0)}_{\alpha\beta\mu_1\mu_2,j}$. Adding the respective energy contribution from each of them yields
\begin{eqnarray}\label{d3_om1_11}
	\Omega_{1,1}^{(0)} &=& \frac{\Delta_{sd}^4}{2}\int_0^\infty \frac{dk k}{4\pi^2}
	\bigg[
	-\frac{\alpha_1^3k^2}{2}(1-m_z^2) r_0^{(2)}(k) L_{jz,j}
	+ \frac{3\alpha_1^2\alpha_2 k^4}{4}r^{(2)}_0(k)
	[2m_x m_y L_{yx,x}+(m_x^2-m_y^2)L_{yx,y}]\notag\\
	&&{}+\frac{3\alpha_1\alpha_2^2 k^6}{2} (1-\frac83m_z^2) r^{(2)}_1(k)L_{jz,j}
	\bigg],
\end{eqnarray}
with the functions $r_{0,1}^{(2)}$ defined as in Eq.~\eqref{r_func}. At this order, it is practical to define three DMI constants to quantitatively describe the free energy,
\begin{equation}
	\Omega_{1,1}^{(0)} = D_{1a}^{(0)}(1-m_z^2)L_{jz,j} 
	+D_{1b}^{(0)}[2m_x m_y L_{yx,x}+(m_x^2-m_y^2)L_{yx,y}]
	+D_{1c}^{(0)}[(1-\frac83m_z^2) L_{jz,j}].
\end{equation}

Finally, there are 80 non-zero components to $D^{(1)}_{\alpha\beta\mu_1\mu_2,j}$. Adding the contributions from each one yields the free energy
\begin{eqnarray}
	\Omega_{1,1}^{(1)}
	&=&\frac{\Delta_{sd}^4}{2}\int \frac{dk k}{(2\pi)^2}\bigg\{\!-\frac{3\alpha_1^5 k^4}{16}r_0^{(3)}
	(1-\frac{m_z^2}{2})L_{jz,j}+\frac{9\a_1^4a_2k^6}{16}
	(r_0^{(3)}-r_1^{(3)})
	[2m_xm_yL_{yx,x}+(m_x^2-m_y^2)L_{yx,y}]\\
	&+&\frac{15\alpha_1^3\alpha_2^2k^8}{16}r_1^{(3)}
	(1-\frac52 m_z^2)L_{jz,j}
	-\frac{9\alpha_1^2\alpha_2^3 k^{10}}{8}r_1^{(3)}[
	2m_xm_y L_{yx,x}+(m_x^2-m_y^2)L_{yx,y}
	]
	+\frac{3\alpha_1\alpha_2^4 k^{12}}{8} r_2^{(3)}(k)m_z^2L_{jz,j}\bigg\}.\notag
\end{eqnarray}
The non-LI invariants that are present in the free-energy expansion to higher order are characterized by qualitatively new invariants of the type $2m_x m_y L_{yx,x}+(m_x^2-m_y^2)L_{yx,y}$. 
These are identical to the non-LI invariant in the $D_{3h}$ group analyzed in Ref.~\cite{Ado2022}. Modulo total derivatives, which vanish in the bulk, they are related as
\begin{equation}
	m_x(m_x^2-3m_y^2)=-\frac{3}{4}(2m_x m_y L_{yx,x}+(m_x^2-m_y^2)L_{yx,y}).
\end{equation}

\paragraph*{Small SOC\\}
Several simplifications are possible in the small SOC limit, where the rotation symmetry breaking SOC distortion to the energy spectrum is treated perturbatively.
The first-order corrections require knowledge of tensors $D_{\alpha\beta,j}^{(0)}$, $D_{\alpha\beta,j}^{(1)}$, and $D^{(0)}_{\alpha\beta\mu_1\mu_2,j}$. In the small SOC limit, at each order one recovers in the integral the rotational symmetry such that the expression for DMI constants is further simplified.

Computed to cubic order in spin-orbit coupling, the nonvanishing components are from~\eqref{D3_D1_sd}
\begin{equation}
	D_{iz,i}^{(0)} \simeq -\frac{\alpha_1\Delta_{sd}^2}{2}\sum_s \int \frac{dkk}{2\pi}
	\bigg[\mathcal F^{(0)}_{s,\bm k}(\Delta_{sd})
	+\frac{k^2}{2}
	\bigg(
	\alpha_1^2 + \frac{\alpha_2^2 k^4}{2}
	\bigg)\mathcal F^{(1)}_{s,\bm k}(\Delta_{sd})
	\bigg].
\end{equation}
This exhibits the same structure as in the $C_{\infty v}$ case. Defining $D^{(0)}_0=D^{(0)}_{xz,x}$, the free-energy density reads
\begin{equation}\label{omega101}
	\Omega_{1,0}^{(0)} = D_0^{(0)}L_{jz,j}.
\end{equation}
Thus working to linear order in SOC yields the conventional LI invariant characterizing the $D_3$ (or $C_{3v}$) point group in 2D (e.g.,~see Ref.~\cite{Bogdanov1989}).

The contribution from $D^{(1)}_{\alpha\beta,j}$ also simplifies since to cubic order in SOC $\mathcal F_{s,\bm k}^{(1)}(\lambda) = \mathcal F_{s,\bm k}^{(1)}(\Delta_{sd})$ and the angular integral is trivial.
Therefore, it readily follows from Eq.~\eqref{d3_om2_10} that
\begin{equation}
	\Omega^{(1)}_{1,0}=D^{(1)}_0 L_{jz,j},
	\quad 
	D_0^{(1)}=-\frac{\alpha_1\Delta_{sd}^2}{16\pi}\sum_s\int dkk^3\big(
	\alpha_1^2-\frac{\alpha_2^2k^4}{2}\big)\mathcal F_{s,\bm k}^{(1)}(\Delta_{sd}),
\end{equation}
which renormalizes the previous term~\eqref{omega101}.

Finally, the last term to cubic order in SOC is the contribution from $D^{(0)}_{\alpha\beta\mu_1\mu_2,j}$.
From Eq.~\eqref{d3_om1_11} it follows directly that the three DMI constants are determined after performing the angular integral in $r^{(2)}_n$ functions,
\begin{eqnarray}\label{3ofD}
D^{(0)}_{1a}&=& -\frac{\alpha_1^3\Delta_{sd}^4}{8\pi}\sum_s\int dk k^3\mathcal F^{(2)}_{s,\bm k}(\Delta_{sd}),
\notag\\
D^{(0)}_{1b}&=& \frac{3\alpha_1^2\alpha_2\Delta_{sd}^4}{16\pi}\sum_s\int dk k^5\mathcal F^{(2)}_{s,\bm k}(\Delta_{sd}), 
\\
D^{(0)}_{1c}&=& \frac{3\alpha_1\alpha_2^2\Delta_{sd}^4}{16\pi}\sum_s\int dkk^7\mathcal F^{(2)}_{s,\bm k}(\Delta_{sd}).
\notag
\end{eqnarray}

In the zero-temperature approximation the DMI constants reveal that there is a nonvanishing contribution due to cubic terms in momentum.
Since both $D_0^{(0)}$ and $D_0^{(1)}$ contribute to the conventional LI, we add them to yield $D_0^{(+)}=D_0^{(0)}+D_0^{(1)}$,
\begin{equation}
	D_0^{(+)} = -\frac{\alpha_1\Delta_{sd}^2}{4\pi}\sum_s\int dkk
	\big[
	F^{(0)}_{s,\bm k}(\Delta_{sd})
	+\frac{k^2}{4}\big(3\alpha_1^2
	+\frac{\alpha^2_2k^4}{2}
	\big)F^{(1)}_{s,\bm k}(\Delta_{sd})
	\big],
\end{equation}
where the $k^6$ term contributes at zero-temperature above the gap,
\begin{eqnarray}
	D_0^{(+)} &=& \bigg[\frac{\alpha_1\Delta_{sd} m}{8\pi\hbar^2}(1-\frac{\mu^2}{\Delta_{sd}^2})
	-\frac{3\alpha_1^3}{16\pi}\frac{\mu}{\Delta_{sd}}\left(\frac{m}{\hbar^2}\right)^2(1-\frac{\mu^2}{\Delta_{sd}^2})
	+\frac{\alpha_1\alpha_2^2}{80\pi}\left(\frac{m}{\hbar^2}\right)^4
	(1+\frac{\mu}{\Delta_{sd}})^3(8-9\frac{\mu}{\Delta_{sd}}+3\frac{\mu^2}{\Delta_{sd}^2})\bigg]
	\Theta(1-\frac{\mu^2}{\Delta_{sd}^2})\notag\\
	&&{}+\frac{\alpha_1\alpha_2^2\Delta_{sd}^2}{5\pi}\left(\frac{m}{\hbar^2}\right)^4\Theta(\mu-\Delta_{sd}).
\end{eqnarray}
Finally, from Eqs.~\eqref{3ofD} we obtain
\begin{eqnarray}
	D_{1a}^{(0)} &=& 
	\frac{3\alpha_1^3\mu}{16\pi\Delta_{sd}}\bigg(\frac{m}{\hbar^2}\bigg)^{\!2}
	\bigg(1-\frac{5\mu^2}{3\Delta^2_{sd}}\bigg)
	\Theta\bigg(1-\frac{\mu^2}{\Delta_{sd}^2}\bigg),
	\notag\\
	D_{1b}^{(0)} &=& \frac{3\alpha_1^2\alpha_2\Delta_{sd}}{32\pi}
	\bigg(\frac{m}{\hbar^2}\bigg)^{\!3}
	\bigg(1-\frac{\mu^2}{\Delta_{sd}^2}\bigg)
	\bigg(1-\frac{5\mu^2}{\Delta_{sd}^2}\bigg)
	\Theta\bigg(1-\frac{\mu^2}{\Delta_{sd}^2}\bigg),
	\\
	D_{1c}^{(0)} &=& \frac{9\alpha_1\alpha_2^2\mu\Delta_{sd}}{16\pi}\bigg(\frac{m}{\hbar^2}\bigg)^{\!4}
	{\bigg(1-\frac{\mu^2}{\Delta_{sd}^2}\bigg)}^{\!2}
	\Theta\bigg(1-\frac{\mu^2}{\Delta_{sd}^2}\bigg).
	\notag
\end{eqnarray}

The SOC in $\mathit{\Gamma}_5$ and $\mathit{\Gamma}_6$ bands in point group $D_3$ reads (see Tab.~\ref{tab:LI})
\begin{equation}
	\Delta_{\rm so}\bm\gamma =(i\alpha_1(k_+^3-k_-^3), \alpha_2(k_+^3+k_-^3), i\alpha_3(k_+^3-k_-^3)),
\end{equation}
with $k_\pm=k_x\pm i k_y$.
Working at small SOC, the integrals are expanded term by term, and we find zero contribution to $\mathcal O(\alpha_i^5)$.

\subsection{DMI constants for LI invariants in all 2D symmetry groups}\label{sec:LI}

\begin{table}[ht]
	\caption{Lifshitz invariants for all the two-dimensional groups obtained to linear order in SOC as contained in the free-energy density $\Omega_{1,0}^{(0)}$. 
		The parameters $\alpha_i$ are real and $\beta_i$ are complex, $k_\pm = k_x\pm i k_y$. 
		The spin-orbit interaction $\Delta_{\rm so}\bm\gamma\cdot\bm \sigma$ is determined by the vector $\Delta_{\rm so}\bm \gamma$ as derived in Ref.~\cite{Samokhin2022} (here modulo an eventual overall sign change).}
	\begin{tabular}{cccc}
		\hline\hline
		Group & $\mathit{\Gamma}$ & $\Delta_{\rm so}\bm\gamma$ & $\Omega_{1,0}^{(0)}$\\
		\hline
		$C_1$ & $\mathit{\Gamma}_2$ & $(\a_1 k_x + \a_2 k_y, \a_3 k_x + \a_4 k_y, \a_5 k_x + \a_6 k_y)$
		& $I_1L_{yz,x}+I_2L_{yz,y}+I_3L_{zx,x}+I_4 L_{zx,y}$
		\\
		& & & ${}+I_5L_{xy,x}+I_6L_{xy,y}$\\
		\hline
		$C_2$ & $\mathit{\Gamma}_{3,4}$ &  $(\a_1 k_x + \a_2 k_y, \a_3 k_x + \a_4 k_y, 0)$ 
		&  $I_1L_{yz,x}+I_2L_{yz,y}+I_3L_{zx,x}+I_4 L_{zx,y}$
		\\
		\hline
		$C_3$ 
		& $\mathit{\Gamma}_{4,5}$ & $(\alpha_1k_x+\alpha_2k_y,-\alpha_2k_x+\alpha_1 k_y,
		\beta_1 k_+^3+\beta^*_1k_-^3)$
		& $I_1(L_{yz,x} + L_{zx,y})+I_2(L_{xz,x}+L_{yz,y})$  \\
		& $\mathit{\Gamma}_{6}$ & $(\b_1 k^3_+ +\b_1^* k^3_-, \b_2 k^3_+ + \b_2^* k^3_-,\b_3 k^3_+ +\b_3^* k^3_-)$ & 0 \\
		\hline
		$C_4$ & $\mathit{\Gamma}_{5,6,7,8}$ & $(\a_1 k_x + \a_2 k_y, -\a_2 k_x + \a_1 k_y, 0)$ 
		& $I_1(L_{yz,x}+L_{zx,y})
		+I_2(L_{xz,x}+L_{yz,y})$ \\
		\hline
		$C_6$
		& $\mathit{\Gamma}_{7,8,9,10}$ & $(\a_1 k_x + \a_2 k_y, -\a_2 k_x + \a_1 k_y, 0)$ 
		& $I_1(L_{yz,x}+L_{zx,y})
		+I_2(L_{xz,x}+L_{yz,y})$ \\
		& $\mathit{\Gamma}_{11,12}$ & $(\b_1 k^3_+ +\b_1^* k^3_-, \b_2 k^3_+ + \b_2^* k^3_-,0)$ &  0 \\
		\hline
		$D_1$ & $\mathit{\Gamma}_{3,4}$ & $(\a_1 k_y, \a_2 k_x, \a_3 k_y)$ & $I_1L_{yz,y} + I_2L_{zx,x} + I_3 L_{xy,y}$ \\
		\hline
		$D_2$ & $\mathit{\Gamma}_5$ & $(\a_1 k_y, \a_2 k_x, 0)$ & $I_1L_{yz,y} + I_2L_{zx,x}$  \\
		\hline
		$D_3$ 
		& $\mathit{\Gamma}_4$ & $(-\alpha_1k_y,\alpha_1k_x,-i\alpha_2(k_+^3-k_-^3))$ 
		& $I_1(L_{zx,x}+L_{zy,y})$ \\
		& $\mathit{\Gamma}_{5,6}$ & 
		$(i\a_1(k_+^3-k_-^3), \a_2(k_+^3+k_-^3), i\a_3(k_+^3-k_-^3))$ & 0  \\
		\hline
		$D_4$ & $\mathit{\Gamma}_{6,7}$ & $(\a_1 k_y, -\a_1 k_x, 0)$ 
		& $I_1(L_{xz,x} + L_{yz,y})$ \\
		\hline
		$D_6$ 
		& $\mathit{\Gamma}_{7,8}$ & $(\a_1 k_y, -\a_1 k_x, 0)$ & $I_1(L_{xz,x} + L_{yz,y})$ \\
		& $\mathit{\Gamma}_{9}$ & 
		$(i\a_1(k_+^3-k_-^3), \a_2(k_+^3+k_-^3),0)$ & 0 \\
		\hline\hline
	\end{tabular}\label{tab:LI}
\end{table}

Here we determine the DMI constants for LIs obtained in the approximation of relatively weak SOC $\Delta_{\rm so}\ll\Delta_{sd}$, extracted from $\Omega_{1,0}^{(0)}$.
The results are presented in Table~\ref{tab:LI} for all symmetry groups.
Generically we see that cubic terms in momentum are irrelevant to first order in spin-orbit coupling. 
This readily yields the DMI constants determined by a single integral,
\begin{equation}\label{dmi_const}
	I_i = \frac{\alpha_i\Delta_{sd}^2}{4\pi}\sum_s\int dk k \mathcal F^{(0)}_{s,\bm k}(\Delta_{sd}).
\end{equation}
In the zero-temperature limit it reads
\begin{equation}
	I_i = - \frac{\alpha_i m\Delta_{sd}}{8\pi\hbar^2}
	\left(1-\frac{\mu^2}{\Delta_{sd}^2}\right)
	\Theta\big(1-\frac{\mu^2}{\Delta_{sd}^2}\big).
\end{equation}
The $I_i$ coefficients are formally the same (up the value of $\alpha_i$) with the one analyzed in detail the $C_{\infty v}$ case, i.e., $D_0^{(0)}$ from Eq.~\eqref{D01}.

\subsection{Gapped Dirac model}
An important limit with application to topological materials is that of a Dirac model with Rashba spin-orbit interactions and gapped by the exchange coupling. 
The Hamiltonian for a $C_{\infty v}$ model reads
\begin{equation}
	H = -\mu\sigma_0 +\alpha_R (\bm k\times\bm \sigma)_z + \Delta_{sd} \bm m(\bm r)\cdot \bm\sigma.
\end{equation}
In this case, the first DMI constants are computed exactly to all orders at zero temperature.

The first DMI coefficient from Eq.~\eqref{D01} in the zero-temperature limit is 
\begin{equation}\label{D01_norm}
	D_0^{(0)}
	=\begin{cases}
		-\frac{\Delta_{sd}\mu}{4\pi\alpha_R},& \mu\in(-\Delta_{sd},\Delta_{sd}), \\
		-\frac{\Delta_{sd}^2}{4\pi\alpha_R}\text{sign}(\mu), & \mu\notin (-\Delta_{sd},\Delta_{sd}).
	\end{cases}
\end{equation}
Note that the zero-order perturbation theory in small spin-orbit coupling would be divergent due to flat bands for the zero-order energy at $\pm \Delta_{sd}$. 
Nevertheless, summation of all orders gives a dispersion to the bands, which returns a finite DMI constant.

To the same order in magnetic texture, $D_0^{(2)}$ from Eq.~\eqref{D02} is half the amplitude of $D_0^{(1)}$, such that the total contribution reads
\begin{equation}
	D_0^{(1)} = -\frac12 D_0^{(0)}, \quad D_0^{(0)}+D_0^{(1)} = \frac12 D_0^{(0)}.
\end{equation}
The first non-Lifshitz invariant, in the zero temperature approximation, is non-zero only in the gap and the related DMI constant from Eq.~\eqref{D11} reads
\begin{equation}
	D_1^{(0)} =  - \frac{\mu\Delta_{sd}}{8\pi\alpha_R}\Theta(1-\mu^2/\Delta_{sd}^2).
\end{equation}
The remaining contribution to $\mathcal O(m^4)$ reads from Eq.~\eqref{D11}
\begin{equation}
	D_1^{(1)} = -\frac{1}{2}D_1^{(0)}.
\end{equation}

\twocolumngrid
\bibliography{bibl}
\bibliographystyle{apsrev4-2}
\end{document}